\documentclass[10pt, conference, letterpaper]{IEEEtran}
\IEEEoverridecommandlockouts

\usepackage{cite}
\usepackage{amsmath,amssymb,amsfonts}
\usepackage{algorithmic}
\usepackage{graphicx}
\usepackage{textcomp}
\usepackage{xcolor}
\usepackage{soul} 
\usepackage[most]{tcolorbox}
\usepackage{multicol}

\usepackage[utf8]{inputenc}

\usepackage{xspace}
\usepackage[square,sort&compress,comma,numbers]{natbib}
\usepackage{balance}
\usepackage{comment}
\usepackage[hyphens]{url}
\usepackage{multirow}
\usepackage{adjustbox}
\usepackage{bm}
\usepackage{makecell}
\usepackage[inline]{enumitem}

\usepackage{todonotes}
\usepackage{subfig}

\DeclareMathOperator*{\ARGMIN}{arg\,min}

\newcommand{\norm}[1]{\left\lVert#1\right\rVert}

\newcommand{\ICARL}{$\mathtt{iCarl}$\xspace}
\newcommand{\MIRAGE}{$\mathtt{MIRAGE}$-2019\xspace}
\newcommand{\OURS}{$\mathtt{iCarl+}$\xspace}

\begin{document}

\bstctlcite{IEEEexample:BSTcontrol} 
\title{A First Look at Class Incremental Learning in\\Deep Learning Mobile Traffic Classification
}

\author{
\IEEEauthorblockN{Giampaolo Bovenzi$^\dagger$, Lixuan Yang, Alessandro Finamore,\\Giuseppe Aceto$^\dagger$, Domenico Ciuonzo$^\dagger$, Antonio Pescap\'{e}$^\dagger$, Dario Rossi}
\IEEEauthorblockA{Huawei Technology France, $\dagger$ University of Napoli Federico II}
}

\maketitle

\begin{abstract}

The recent popularity growth of Deep Learning~(DL) re-ignited the interest towards traffic classification, with several studies demonstrating the accuracy of DL-based classifiers to identify Internet applications' traffic. Even with the aid of hardware accelerators (GPUs, TPUs), DL model training remains expensive, and limits the ability to operate frequent model updates necessary to fit to the ever evolving nature of Internet traffic, and mobile traffic in particular.
To address this pain point, in this work we explore \emph{Incremental Learning} (IL) techniques to add new classes to models without a full retraining, hence speeding up model's updates cycle.
We consider \ICARL, a state of the art IL method, and \MIRAGE, a public dataset with traffic from $40$ Android apps, aiming to understand \emph{if there is a case for incremental learning in traffic classification}. By dissecting \ICARL internals, we discuss ways to improve its design, contributing a revised version, namely \OURS. Despite our analysis reveals their infancy, IL techniques are a promising research area on the roadmap towards automated DL-based traffic analysis systems.

\end{abstract}

\section{Introduction}

Traffic classification (TC) is at the core of any network traffic monitoring system, and a pillar for traffic management, cyber security, quality-of-experience monitoring, and other strategic activities for network operators. It is also a very mature research topic with many surveys on the subject~\cite{tcsurvey08comst,tcsurvey15survey,tcsurvey18comst}.

From a chronological standpoint, we can categorize TC literature into two ``waves''.
The first wave ignited in the early 2000's, and centered around the use of Machine Learning (ML) methods using per-packet (e.g., packet size, packets inter-arrival time) or per-flow (e.g., total bytes, packets, ports) features as input targeting the classification of a handful of applications. Several works demonstrated that even when just a few packets of a flow were observed, the classification was accurate~\cite{bernaille06ccr,crotti07ccr}, and could be sustained at line-rate speed~\cite{santiago12imc}---``early'' TC was born.

Inspired by the success of image processing in computer vision, in the last years Deep Learning (DL) techniques re-ignited the interest towards TC, with several DL-based classifiers being proposed using as input either raw payload bytes or the same traffic features discovered during the first wave~\cite{wang2015blackhat,wang2017isi,aceto2018tma,lopez2017access,aceto2019mirage,lotfollahi2020deep}.
This renewed interest stems also from the growth in adoption of traffic encryption, the extreme dynamicity of Internet traffic, and the heterogeneity of devices connecting to the Internet, especially when considering mobile ones (having an ecosystems of tools that eases the installation of new apps and their updates)~\cite{aceto2019mirage}.

\begin{figure}[!t]
\centering
\includegraphics[width=0.65\columnwidth]{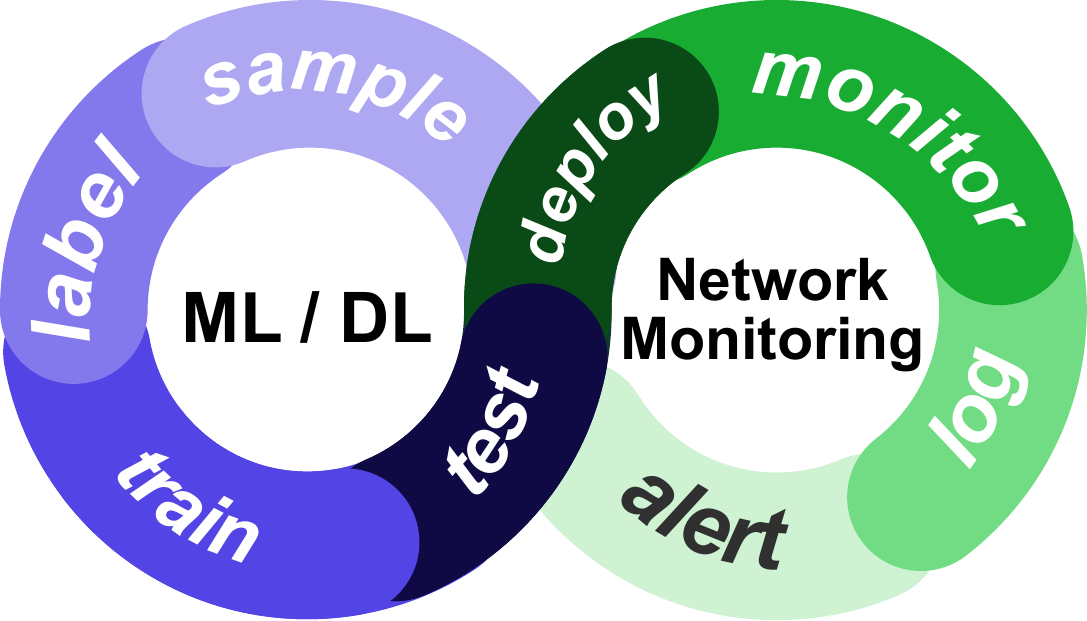}
    \caption{Models development infinite loop.}
    \label{fig:dataflywheel}
\end{figure}

Accordingly, to closely track the network traffic landscape, an effective TC system should support \emph{continuous} model updates, as sketched in Fig.~\ref{fig:dataflywheel}. 
\emph{Incremental Learning} (IL), also known as \emph{continuous} or \emph{online} learning \cite{Van2019}, is a discipline studying 
how to update models to accommodate the new knowledge required to perform well the target task (e.g., a new class needs to be added to a classifier).
This fits well TC needs, and a few works indeed consider incremental TC. However, those works resort to datasets with only a few classes (typically less than $10$), they use per-flow features (which only enable post-mortem classification~\cite{hat-TELECOMSYST16,isvm-MNA18}), and they do not consider scenarios where new applications are progressively added to models (needed to perform network management on new services or apps). In other words, TC literature focuses only on the problem of creating the most accurate classifier given a dataset where both ($a$) the number of classes and ($b$) the data for each class are immutable. 
In turn, TC systems based on literature approaches use \emph{amend and retrain} policies: to add new applications or new traffic behavior to a model one needs to ($i$) create a new training set (or expand the existing one), and ($ii$) train a new model from scratch---model updates \emph{are not incremental}.

Differently, in this paper we investigate the use of IL techniques~\cite{Van2019} to expand the knowledge of an existing DL-based traffic classifier without requiring a full retraining, namely \emph{Class Incremental Learning} (CIL), which we explore to support real-world operations by reducing the requirement of full models retraining. 
We apply \ICARL~\cite{icarl-CVPR17}, a CIL state-of-the-art approach, to a standard 1-dimensional convolutional neural network TC model, using the publicly available \MIRAGE dataset comprising traffic of $40$ popular Android applications. Our contributions are two-fold: ($i$) we dissect \ICARL's design and propose alternative options, including a revised version of the original method we named \OURS, and ($ii$) a thorough evaluation highlighting relevant trends, pitfalls, and further directions of improvement.

The rest of the paper is organized as follows:
in Sec.~\ref{sec:background} we review IL considering both ML and DL methodologies, and the relevant TC literature; in
Sec.~\ref{sec:methodology} we drill down into \ICARL's design, and set our research questions which we evaluate in Sec.~\ref{sec:results}; finally, in
Sec.~\ref{sec:conclusions} we conclude pointing to future avenues of research.

\begin{figure}[t]
    \centering
    \includegraphics[width=\columnwidth]{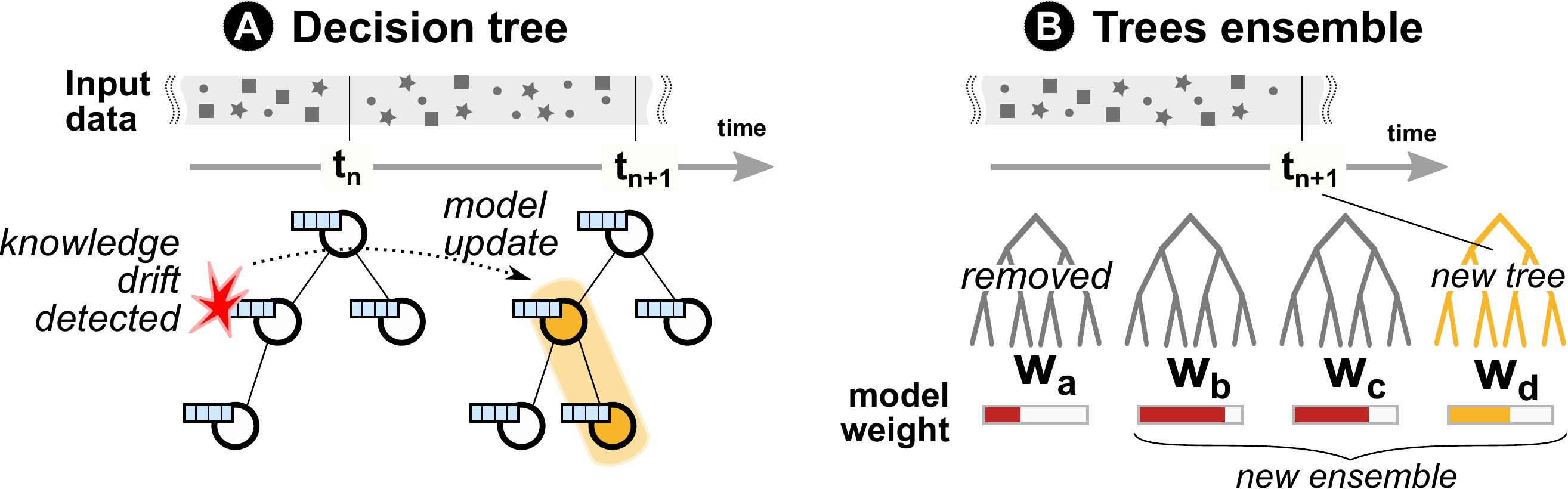}
    \caption{Incremental learning in tree-based classifiers.}
    \label{fig:sketch-ml}
\end{figure}

\section{Background\label{sec:background}}

Incremental learning studies how to ($i$) integrate new classes and ($ii$) refresh the knowledge of already known classes without retraining a model from scratch. Models update procedures are designed to limit \emph{catastrophic forgetting}, i.e. the model ``forgets'' the knowledge already acquired in favor of the new one~\cite{catastrophic-PSM89,catastrophic-ICLR14}, and they rely also on a \emph{memory} accumulating input samples in between updates. Herein, first we review how ML- and DL-based IL techniques (typically) incorporate these two principles, and then we introduce our research goals.

\subsection{Machine learning approaches}

Incremental learning in ML is defined in the context of processing continuous streams of data, i.e., scenarios where is unfeasible to adopt large memories, and operate multiple scans of the data for an update.
In these cases, tree-based data structures offer the best performance, apt to integrate knowledge carried in either individual or batch of data~\cite{batchvsincremental-IDA12} with well established algorithms. 

For instance, in a Hoeffding decision tree~\cite{hat-IDA09} each node buffers statistics observed in a window of data; ADWIN~\cite{adwin-SIAM07} then detects knowledge drift in the buffered data, and an update takes place for example by branching out new nodes (Fig.~\ref{fig:sketch-ml}-A). In a trees ensemble~\cite{Parker2015} instead, each tree is weighted by its accuracy and, using algorithms such as AWE~\cite{awe-KDD03}, trees with low weights are pruned from the ensemble (Fig.~\ref{fig:sketch-ml}-B). 

We underline that these algorithms, and other ML alternative techniques, update only the knowledge of already known classes, and do not incrementally add new ones. This justifies the scarcity of TC literature exploring them, since the typical dataset is a ``static snapshot''. For instance, Hoeffding Adaptive Trees are contrasted against decision trees in~\cite{hat-TELECOMSYST16} to identify 10 classes of traffic evolving over 12 years of MAWI traces. We did not find any work adopting incremental trees ensembles, while a few works use incremental support vectors~\cite{isvm-MNA18} and $k$-means~\cite{inckmeans-ACISC16} although in a more constrained scenario than~\cite{hat-TELECOMSYST16}.  

\begin{figure}[t]
    \centering
    \includegraphics[width=\columnwidth]{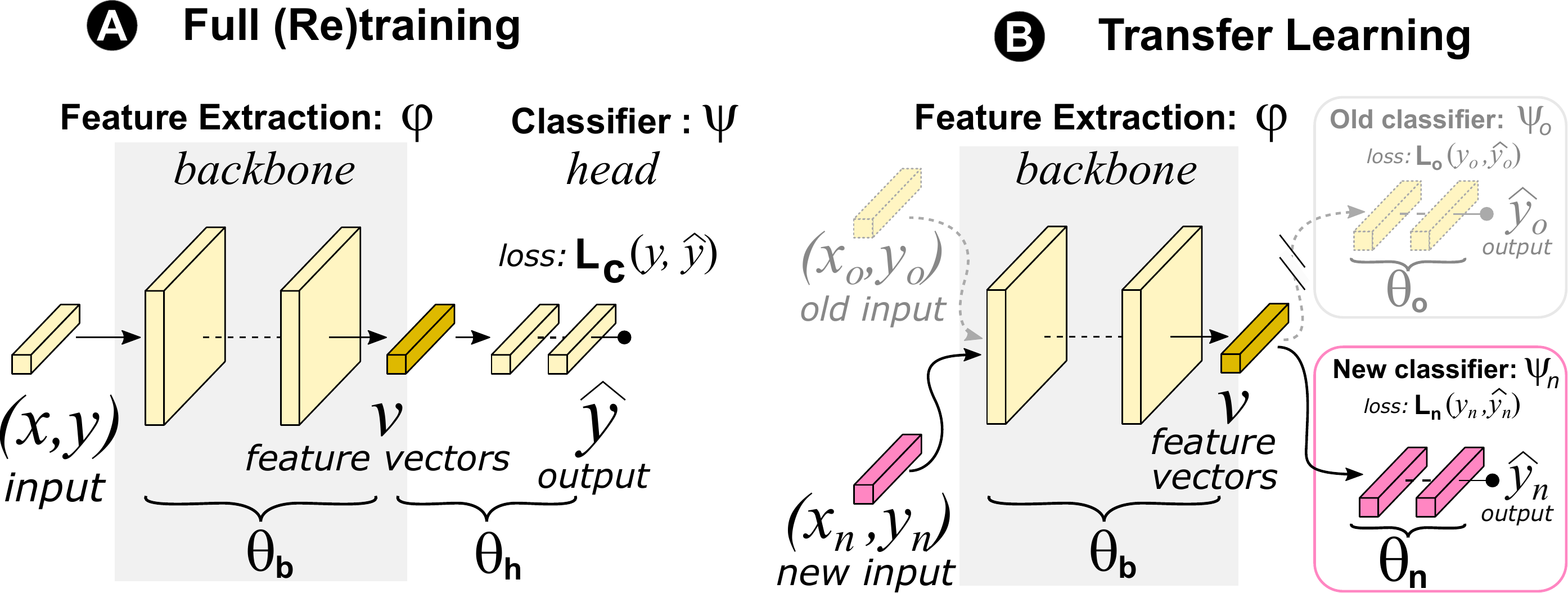}
    \caption{Deep learning model internals (A) and transfer learning application (B).}
    \label{fig:sketch-dl-base-transferlearning}
\end{figure}

\subsection{Deep learning approaches\label{sec:background-dl}}

DL models are harder to evolve than trees. Yet, their ``modularity'' enables a better reuse of knowledge extraction than trees, and opens new venues for IL.
 
To better understand this, Fig.~\ref{fig:sketch-dl-base-transferlearning}-A sketches the internals of a DL model: a composition of \emph{feature extraction} and \emph{classification}, each corresponding to a separate data transformation.
First, $\varphi : X$$\rightarrow$$V$ transforms an input sample $x$ into a feature vector $v$. Then, $\psi : V$$\rightarrow$$Y$ splits the feature vectors $v$ based on their true label $y$, obtaining the final classification $\hat{y}$. The function $\varphi(\cdot)$ and its parameters $\theta_b$ are known as the \emph{model's backbone}; $\psi(\cdot)$ and its parameters $\theta_h$ are the \emph{model's head}. The rationale of this composition is that feature extraction $\varphi(\cdot)$ aims to increase the input space dimensionality so to bring closer points sharing the same label while distancing points of different labels, and to ease the formulation of the classification task $\psi(\cdot)$. 
At training, feature extraction and classification are performed together to discover the best combinations of parameters $(\theta_b, \theta_h)$ minimizing a loss function $L_c(y,\hat{y})$ which compares the true label $y$ and the inferred one $\hat{y}$. The $(\theta_b, \theta_h)$ pair embodies the model knowledge, so adding new knowledge requires changing those parameters and/or the model architecture.

Given this model composition, the IL literature broadly falls into \emph{three} categories: \emph{model-growth} approaches enlarge models architecture and layers size to accommodate for new knowledge~\cite{mallya2018, aljundi2016};  \emph{fixed-representation} approaches rely on a rich backbone~\cite{kemker2018, hayes2020} while altering the model's head; and \emph{fine-tuning} adapt gradually model's backbone and head~\cite{lwf-TPAMI18, icarl-CVPR17}.
All those methods try to address the limited capability of neural networks to support incremental changes, thus being significantly more sensible to catastrophic forgetting than trees.
To address this issue, incremental learning methods typically rely on  \emph{Knowledge distillation}~\cite{distillation-NIPS15}, a technique that decomposes the training loss function $L$ into two terms: a \emph{distillation} term $L_o$ quantifying the (old) knowledge acquired, and a \emph{classification term} $L_n$ quantifying the (new) knowledge to add.
The two terms can be balanced in different ways~\cite{lwf-TPAMI18}. 

Moreover, differently from ML techniques, memory plays a key role since
($i$) training requires multiple passes on the input data and
($ii$) \emph{rehearsal mechanisms}, i.e., re-proposing samples from a previous training set when updating the model, are important to overcome catastrophic forgetting~\cite{Van2019,belouadah2020} and avoid the extra cost of data augmentation techniques like generative replay~\cite{Shin2017}.

\noindent\textbf{Transfer learning.}

A model's backbone learns hidden patterns in the data, and the larger the training set (and the lower the layer in the model's architecture), the more the extracted knowledge is expected to abstract from the specific classification task at hand.
This observation is key in \emph{transfer learning} where the backbone of a model $M_a$ is reused for a model $M_b$ by detaching $M_a$ head $\theta_o$ and replacing it with a new one $\theta_n$ tailored to the new classification task (Fig.~\ref{fig:sketch-dl-base-transferlearning}-B). Then, $M_b$ is trained via fixed-representation or fine-tuning.
Considering TC, the authors of~\cite{multitasklearn-ICCCN20} show that one can reuse the backbone of a model trained to predict flows duration and bandwidth, to train a classifier to distinguish 5 traffic classes. 

In its basic formulation, transfer learning derives a new model from an existing one, hence does not perform an update. It is however also possible to incrementally add new tasks to an existing model~\cite{neverending-COMM18}---this is known as \emph{multi-task incremental learning}. In this case, a new head is attached to the backbone without pruning the existing ones. 
Considering traffic classification, in~\cite{fewshot-IEEEAccess19} authors study how to add a traffic classifier task to a model trained to classify both flow duration and flow bandwidth.

\begin{figure}[t]
    \centering
    \includegraphics[width=\columnwidth]{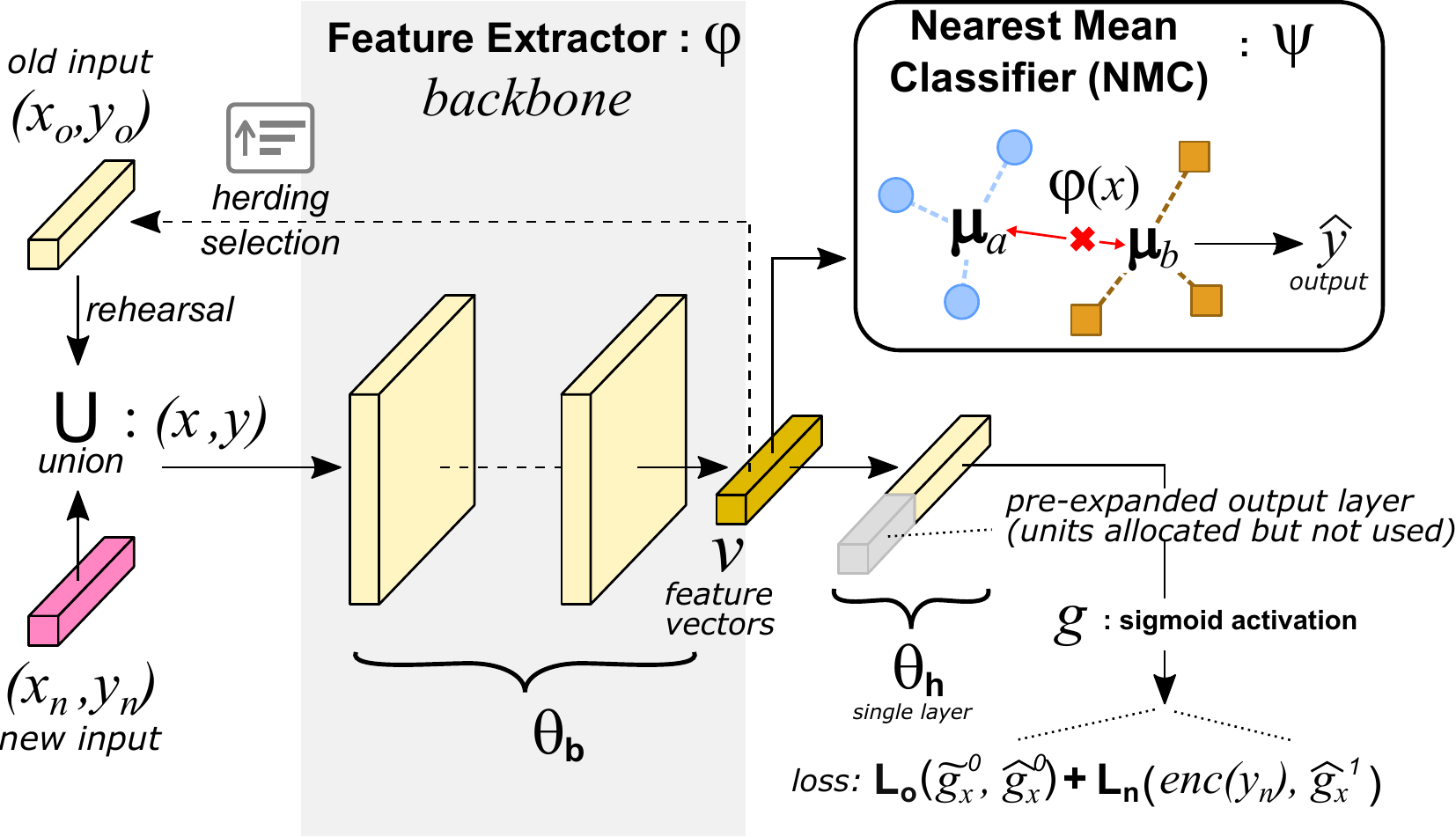}
    \caption{Sketch of \ICARL~\cite{icarl-CVPR17} internals.}
    \label{fig:sketch-icarl}
\end{figure}

\noindent\textbf{Class incremental learning (CIL).}

If we do not stick to rigid taxonomy boundaries, transfer learning can also be used for CIL. To be more specific, a CIL state-of-the-art method is \emph{Incremental Classifier and Representation Learning} (\ICARL)~\cite{icarl-CVPR17}.
\ICARL is a fine-tuning technique using ($i$) known classes exemplars rehearsal, ($ii$) a pre-allocated output layer, ($iii$) knowledge distillation, and ($iv$) a Nearest Mean Classifier (NMC), as sketched in Fig.~\ref{fig:sketch-icarl}.

Let us consider a base model $(\theta_b, \theta_h)$ trained on a $(X_o, Y_o)$ dataset and having $C$ classes. At the input, \ICARL selects exemplars of known classes $P$$\subseteq$$X_o$ which are rehearsed at training.
A herding mechanism~\cite{herding-ICML09} selects exemplars based on their distance with respect to the class average centroid in the latent space, and keeps them in a memory of fixed size.

At the output, the model's head is a single layer (with neurons using the sigmoid activation function) of size $K$ with $K$$\gg$$C$---the head pre-allocates \mbox{$K$-$C$} units for future classes to be added.
We denote with $\hat{g}_x$ the response vector associated to $K$ sigmoids, whereas with ${\hat{g}_{x}^{0}}$ (resp. ${\hat{g}_{x}^{\mathrm{f}}}$) the subvector associated to the base $C$ (resp. future \mbox{$K$-$C$}) classes.
The classification is however operated computing distance from classes' average centroid $\mu_c$ in the latent space \mbox{$\hat{y} = \ARGMIN_i \norm{ \varphi(x) - \mu_c }$}---this is known as \emph{Nearest Mean Classifier} (NMC). 
In computer vision, this simple mechanism has been found superior to more typical parametric classification (i.e., leveraging the model head) when performing incremental learning~\cite{nmc-TPAM13,nmc-CVPR14}.

When performing an update, \ICARL first applies the current model to \emph{both} exemplars in memory and new classes data.
The obtained ($C$-dimensional) soft-output $\tilde{g}_{x}^{0}$ is used for knowledge distillation $L_{o}(\tilde{g}_{x}^{0},\hat{g}_{x}^{0})$.
Additionally, some of the $K$-$C$ extra units pre-allocated in the model's head (i.e. $\hat{g}_{x}^{\mathrm{f}}$), denoted with $\hat{g}_{x}^{1}$, are ``consumed'' for the new classes.
These units are trained via the classification loss, whereas labels of new (resp. base) classes are one-hot (resp. zero-vector) encoded, namely $L_n(enc(y_n), \hat{g}_{x}^{1})$.

\subsection{Our research goal}

We found scarce literature for IL applied to TC. 
Indeed, only \cite{transflearn-INFOCOM20} presents a closed-loop TC system,
but authors ($i$) consider a small use-case with $10$ classes, ($ii$) adopt simple fine-tuning to introduce new classes, and ($iii$) focus on a single model update. Broadly speaking, the closely related literature, no matter the adopted methodology, share common limitations: old datasets (\cite{isvm-MNA18,inckmeans-ACISC16} use data from $2005$, whereas \cite{hat-TELECOMSYST16} use 2001-2014 data); very few (typically less than $10$), and very generic classes (HTTP, DNS, SSH, etc.); adopt per-flow features, thus not performing early traffic classification.
On the one hand, the validation of CIL techniques against outdated datasets allow to grasp insights on models generalization and conjecture on their future longevity. 
On the other hand, some datasets are $10+$ years old, and offer just a handful of classes, which makes them unfit for our study.
Indeed, the limited availability of public labeled datasets for TC is a well known problem,
quite far from the abundance of datasets for computer vision- and natural language processing-related tasks.

Methodology-wise, we aim to a fresh look at CIL for traffic classification.
We discard ML-based IL techniques since they can only expand knowledge of already known classes, and rely on handcrafted features.
We also discard multi-task transfer learning techniques since they lead to multi headed models which complicate systems operation---given one input sample, a multi headed model would return one label for each head, raising the problem of how to reconcile multiple output labels into one.
We focus instead on fine-tuning, and specifically \ICARL given its accuracy, limited complexity, and high scalability~\cite{icarl-CVPR17}.  
To the best of our knowledge, \ICARL has only been applied to image classification, hence we aim to understand if its qualities remain the same when used for (mobile) TC.
To do so, we contribute an in-depth analysis of \ICARL internal design, challenging the original choices (e.g., the adoption of NMC), and thoroughly evaluating alternative choices using the publicly available \MIRAGE which, to the best of our knowledge, is the most recent dataset on mobile traffic, and offer a large variety of classes ($40$ Android apps), thus fitting well the aim of our study.

\section{Methodology}\label{sec:methodology}

To better understand the design space, in this section we start reviewing \ICARL internal components in more details, and set our research questions.
We then introduce the dataset, and the DL-based model we use in our evaluation.

\subsection{Dissecting \ICARL design}

\ICARL is a CIL state-of-the-art approach for image classification 
which trades off complexity and performance with respect to alternative solutions like~\cite{castro2018end, wu2019}, thus is the perfect candidate technique for our initial exploration of CIL for TC.

\ICARL's design centers around two components: a \emph{memory} at the input, and an \emph{NMC} at the output (Fig.~\ref{fig:sketch-icarl}), both operating on the feature vectors generated by a model's backbone. 
The memory is fixed in size, and contains a uniform amount of exemplars for each class.
It follows that, at each update, each known class drops the same amount of exemplars to make room for the new classes exemplars. Such filtering is controlled by a herding mechanism, a procedure aiming to select a subset of exemplars so that the class average centroid in the latent space created from the filtered exemplars is as close as possible to the one obtained without filtering---it selects a ``herd'' of points capturing ``the body'' of the class (for more details, please refer to~\cite{icarl-CVPR17}). 
By design, an NMC tights to the same mechanics of centroids and point-to-centroid distances computation in the latent space. According to~\cite{icarl-CVPR17}, these design choices aim to \emph{decouple the non-linear feature extraction from the linear classifier}: while the NMC approximates the class feature vectors mean, the herding selection process tries to keep this mean unchanged by carefully selecting exemplars---this is the \emph{representation learning} component in the \ICARL acronym, which stresses the need of a robust feature extraction. 

At a closer look, we spot \emph{two} other subtle aspects, at the border between design choices and implementation details. First, \ICARL adopts a sigmoid rather than a softmax activation function at the output. We find this peculiar, and while authors briefly state that both functions are equivalent, they do not evaluate the impact of softmax.
Secondly, as introduced in Sec.~\ref{sec:background-dl}, the output layer is pre-allocated, i.e., its size is larger than the number of classes required, and the extra units are progressively used when adding new classes.
\ICARL authors do not mention this in their paper, but it is evident inspecting \ICARL source code.\footnote{https://github.com/srebuffi/iCaRL} While at first sight this sounds like a ``smart trick'', it can cause unexpected effects when intertwined with the use of sigmoid activations (see Sec.~\ref{sec:results-upperbound}).

Considering the overall design choices,
in this work we target the following questions:
\emph{
\begin{itemize}
    \item Is it better to use an NMC, or to classify via the output layer? 
    \item Is it better to adopt a sigmoid, or a softmax activation function at the output layer?
    \item Is it better to pre-allocate the output layer size, or progressively expand it?
    \item How sensitive is the herding selection to memory size? Does a larger memory provide better performance?
\end{itemize}
}

These are all general questions of practical relevance, and to the best of our knowledge they are not fully investigated in the literature.
Yet, it is beyond the scope of this work to provide a final answer to them. Rather, we explore the questions in the context of TC using only one dataset.
In particular, differently from observed in~\cite{icarl-CVPR17}, we find that ($i$) output layer classification is superior to NMC, ($ii$) softmax is superior to sigmoid, ($iii$) pre-allocation of the output layer may have a detrimental effect, and ($iv$) memory has indeed an impact in balancing catastrophic forgetting as reported in literature. We contribute our observations into a revised version of \ICARL, namely \OURS.
As we shall see in Sec.~\ref{sec:results-il-strategies}, neither \ICARL nor \OURS match the performance observed in computer vision when tested on a TC dataset. Yet, \OURS significantly reduces training time, another aspect not quantified in previous literature (Sec.~\ref{sec:results-training-time}). 

\subsection{Dataset}

\begin{figure}[!t]
    \centering
    \includegraphics[width=\columnwidth]{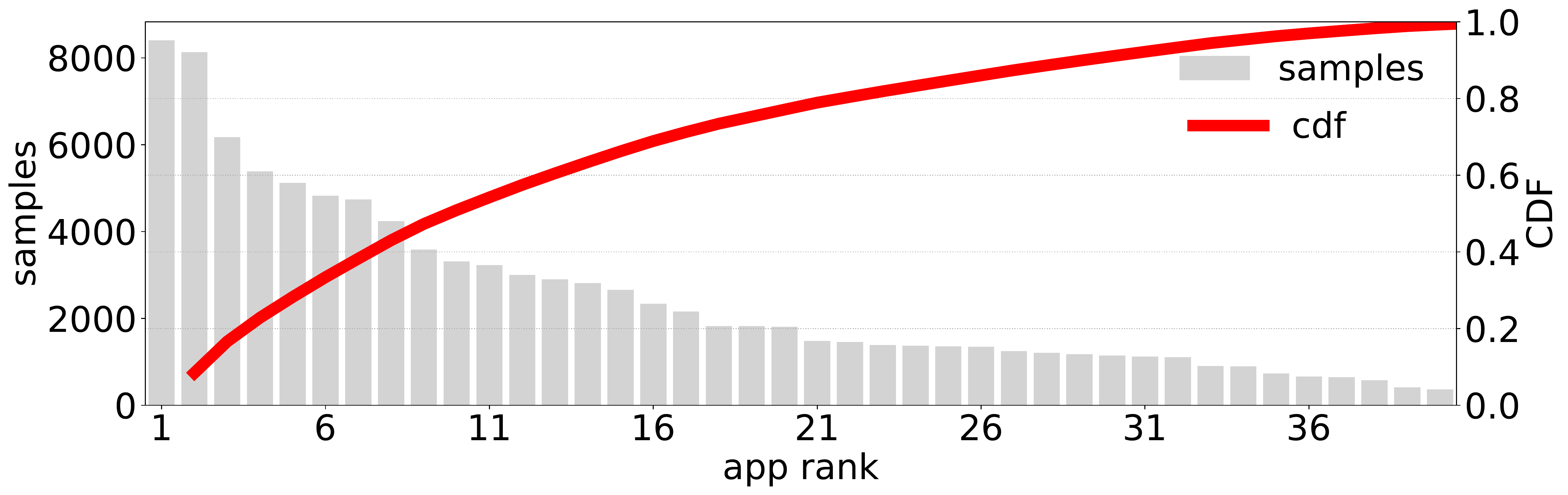}
    \caption{\MIRAGE dataset composition.}
    \label{fig:dataset}
\end{figure}

\begin{figure}[!t]
    \includegraphics[width=\columnwidth]{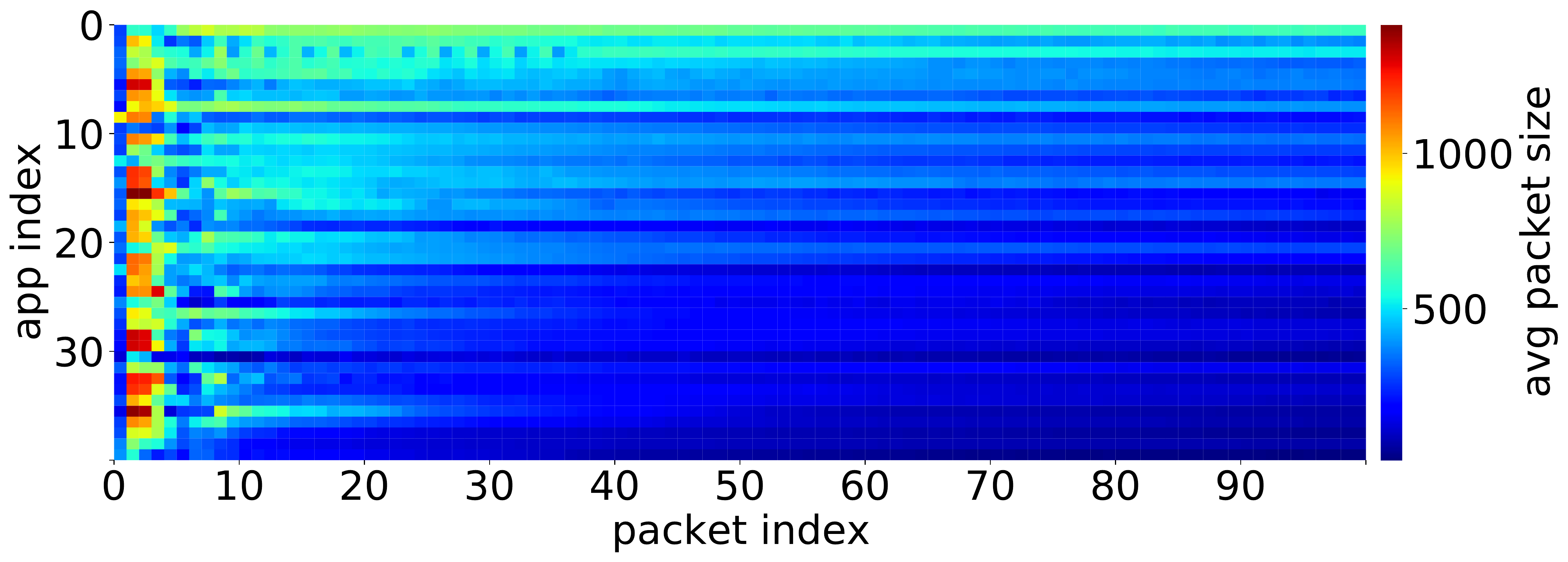}
    \includegraphics[width=\columnwidth]{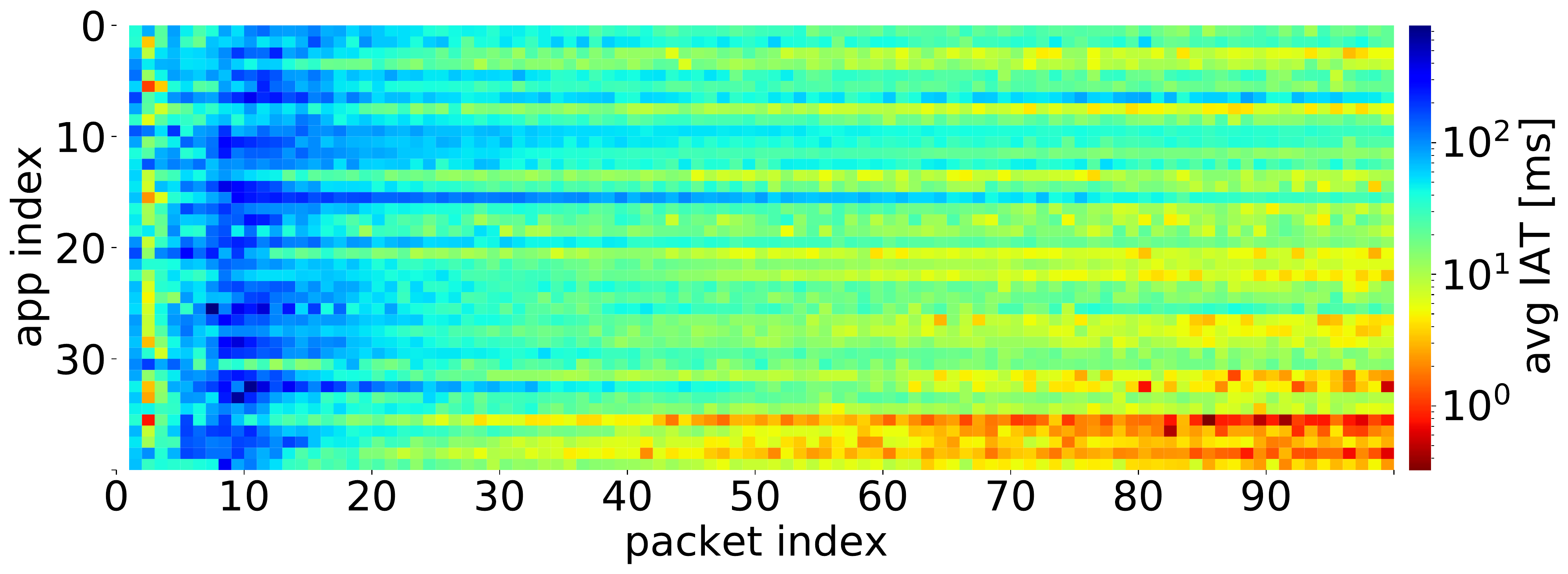}
    \includegraphics[width=\columnwidth]{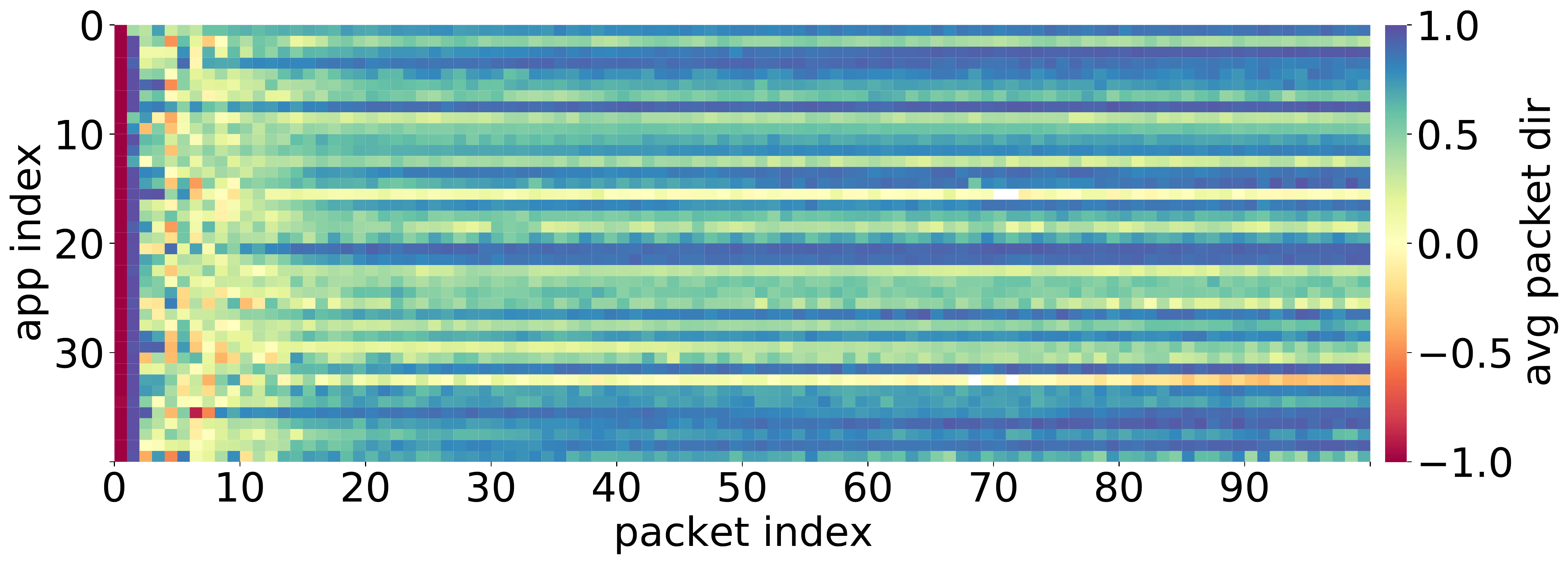}
    \includegraphics[width=\columnwidth]{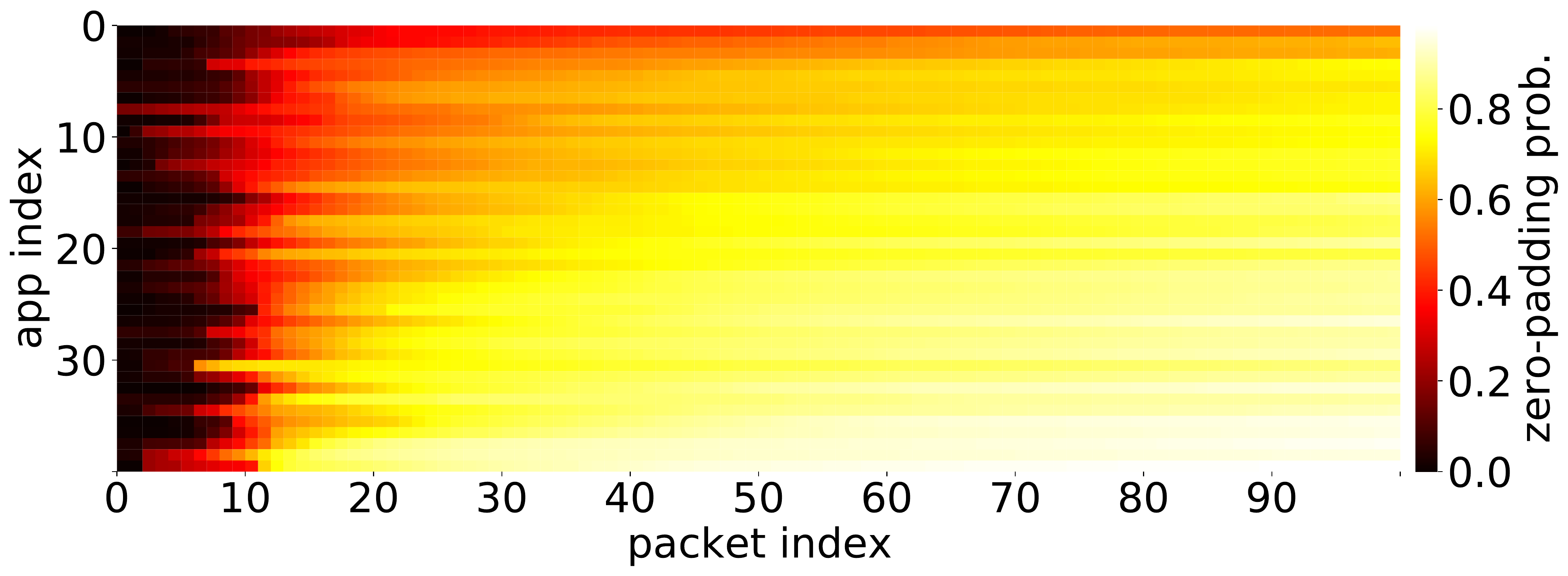}
    \caption{\MIRAGE dataset biflows time series properties.}
    \label{fig:heatmaps}
\end{figure}

In this work we use the \MIRAGE~\cite{aceto2019mirage} dataset.
It contains traffic related to $40$ Android apps belonging to $16$ different categories.\footnote{\url{http://traffic.comics.unina.it/mirage/app_list.html}}
\MIRAGE was collected at the ARCLAB laboratory of the University of Napoli ``Federico II'' with multiple measurement campaigns running between May 2017 and May 2019.
In each experiment a volunteer (out of $\approx$300 students) was asked to use one of the $40$ apps while in background raw pcap files and \texttt{strace} log-files were collected on the phones. Then, pcap were converted into JSON files where each record reports on the network activity of a different bidirectional flows (biflows) with aggregate level metrics (total bytes, packets, etc.), and per-packet level metrics (time series of packet size, direction, TCP flags, etc.).
The \texttt{strace} logs instead exposed the socket-to-appID mapping as observed on the phone, so they were used as ground-truth to label the per-flow logs obtained from raw pcaps. For more details, please refer to~\cite{aceto2019mirage}.

The \MIRAGE dataset contains $\approx$100k biflows distributed across apps as shown in Fig.~\ref{fig:dataset}.
Although the number of experiments for each app was comparable, biflows' distribution is heavy-tailed with the top-10 (top-20) apps accounting for $54.2$\% ($78.9$\%) of all biflows.

In this work, we focus on three biflow-related per-packet properties, namely packet (L4) payload size (PS), inter arrival time (IAT), and packet direction (DIR).
Those have been successfully used in previous literature for early TC~\cite{bernaille06ccr}.
Discarding packets with zero payload, we find that $86\%$ of biflows have size $\leq 100$ packets, while
the average size of biflows across the entire dataset is $180$ packets. Thus, for each flow, we extract time series of length $100$ for PS, IAT, and DIR, that we use an input for our DL classifier.
In case the time series are shorter than 100 packets, we apply zero-padding. Notice that we encode DIR values as $+1$ (for upstream) or -1 (for downstream), while the minimum IAT supported by the \MIRAGE capture is $1\mu$s, so the zero-padding does not clash with valid time series values.

In Fig.~\ref{fig:heatmaps} we render the obtained time series as heatmaps:
each row maps to a different app, and columns map to a different time series positions, with values  averaged across all biflows.
The bottom heatmap further shows the probability of zero-padding, which we also use to sort heatmaps' rows by putting more ``chatty'' apps first.

Considering packets size, all apps have a burst within the first $5$ packets, more sporadic communications within the first $25$ packets, after which they are mostly silent. Both zero-padding and packet direction heatmaps well capture this effect too.
The IAT's heatmap further captures the interleaving period between first and second communications. This interleave is possibly related to the nature of mobile apps which, based on HTTPS, render content in stages~\cite{mcpa-TMA19}.
The right side of the IAT's heatmap also shows bursts of activity 
but, especially for the apps at the bottom of the heatmaps, this is likely an artifact of the reduced number of packets (notice the higher zero-padding for those apps).

\begin{figure}[!t]
    \centering
    \includegraphics[width=\columnwidth]{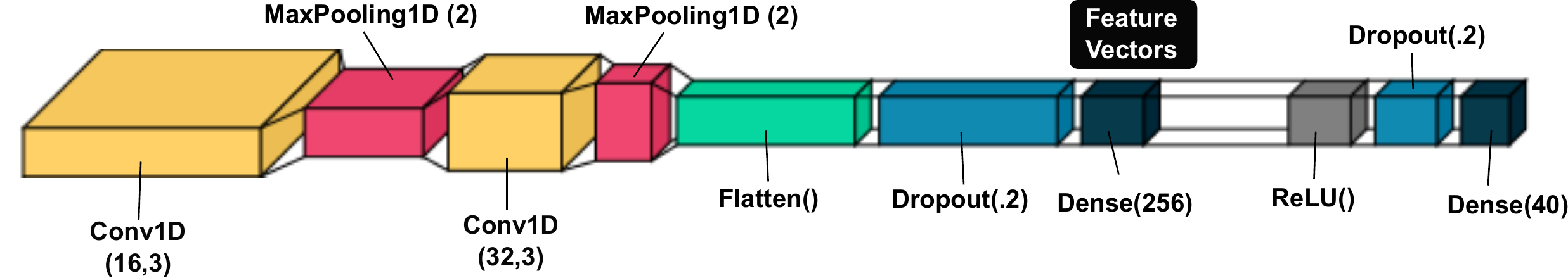}
    \caption{Select DL model architecture.}
    \label{fig:architecture}
\end{figure}

\subsection{Model architecture}\label{sec:model-arch}

In theory, \ICARL can be used without changing a model's architecture, with the caveat that the original design requires to adopt a sigmoid activation as output, and the model head is expected to be a single layer. 
In other words, \ICARL puts an accent on the importance of the model's backbone. \ICARL is originally evaluated using a convolutional neural network (CNN), which well suits TC needs too~\cite{aceto2018tma}. Thus, we rely on the 1d-CNN architecture sketched in Fig.~\ref{fig:architecture}.
The input is a $3$ channels $100$-elements time series where each channel is associated to a different packet property. 
The considered input configuration was observed experimentally to provide higher performance with respect to configurations with ($i$) only (PS,DIR) pairs, and ($ii$) PS*DIR (i.e. signed payload sizes), with $+1\%$ and $+6\%$ F1 score improvement, respectively.
We also experimented with time series with less than 100 packets, and obtained only marginal differences in model accuracy.
The input layer feeds a stack of $2$ convolutional layers, with $16$ and $32$ filters of $1\times3$ size, respectively, ReLU activations and max-pooling layers (stride $2$), followed by one fully-connected layer of size $256$ for a total depth of $3$ layers before the output layer.
This corresponds to $\approx200$k parameters overall. 

We underline that we do not claim any novelty on the model architecture, rather we point out that similar designs have been found accurate in previous literature~\cite{aceto2018tma,wang2017isi}, so we consider it a good reference choice and a state-of-the-art network for TC.
In selecting the architecture we also intentionally avoided including payload bytes or long short term memory (LSTM) layers which could have possibly increased the model performance. Rather, we opted for a ``conservative'' base line.
Still, we highlight that the considered IL methodology virtually applies to other DL-based traffic classifier proposals.

\section{Evaluation\label{sec:results}}

We compared three strategies for CIL, namely \ICARL, fixed representation, and \OURS, with the last two derived from \ICARL's source code. 
We evaluated all strategies updating \emph{upperbound} models
i.e., models trained from scratch using the architecture defined in Sec.~\ref{sec:model-arch}. Upperbound models are the performance baseline for models incrementally trained using \ICARL.
We leveraged the apps variety in the \MIRAGE to construct different scenarios varying the number of base classes, performing individual or multiple updates in sequence, and using two or more classes for each update.
We carried out $10$ runs for each scenario by randomizing the set of classes. Unless differently reported, all runs used a memory of $1\mathrm{k}$ exemplars.
We measured TC effectiveness via (macro average) F1 score, which we split between \emph{base classes} and \emph{new classes} to quantify catastrophic forgetting.

We adopted the same parameters used by \ICARL: all models (both upperbound and updates) were trained for $200$ epochs with a learning rate of $10^{-2}$ halved every $50$ epochs, and a momentum of $0.9$; for model updates only, the loss function had an extra regularization term with $10^{-5}$ weight.

In the remainder, we introduce the selection process of the upperbound, and evaluate the design choices at the output. We then contrast the three IL strategies, and further zoom into \OURS to assess the impact of both multiple updates and memory size. We conclude reporting on model update time.

\begin{figure}[t]
    \centering
    \includegraphics[width=\columnwidth]{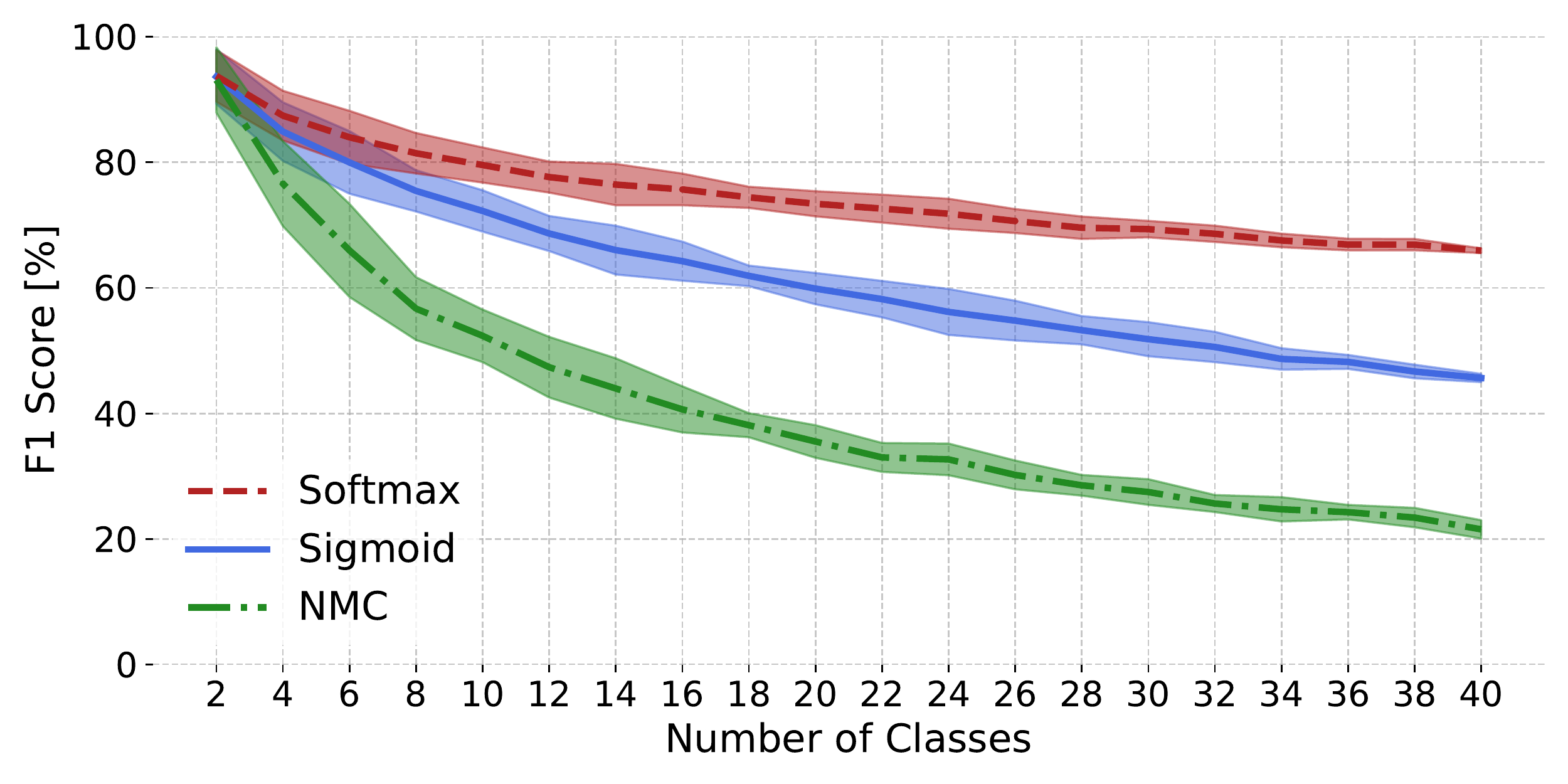}
    \caption{Classification methods accuracy in upperbound models.}
    \label{fig:upperbound}
\end{figure}

\subsection{Design choices at the output\label{sec:results-upperbound}}

We investigated three design choices at the output: ($i$) the type of activation function, ($ii$) the type of classifier, and ($iii$) the method to expand the output layer.
Without loss of generality, we quantified all of them directly using upperbound models. 

Focusing on choices ($i$)-($ii$), Fig.~\ref{fig:upperbound} compares three configurations: upperbound models classifying by means of the output layer via either softmax or sigmoid, and the original \ICARL design, i.e., a sigmoid activation coupled with an NMC. Lines correspond to averages across runs, while shaded areas capture standard deviation.
\ICARL's design offers the worst performance, even worse than classifying via a model's head using a sigmoid activation.
This differs from results in~\cite{icarl-CVPR17}, where \ICARL's authors find an NMC to be superior than classifying via the output layer trained with sigmoid.
The best strategy is also the most popular in literature: to classify via the model's head and adopt a softmax activation (not evaluated in~\cite{icarl-CVPR17}). We conjecture that the NMC poor performances are possibly due to:
($a$) our models being smaller, both in terms of architecture and training set size, than what commonly used in computer vision, hence the feature representation obtained from the model's backbone might not be robust enough to integrate well new classes;
($b$) the lack of an effective decoupling between feature extraction and classification might be biasing NMC class centroids, since they are expected to rely on class-conditional distributions following the multivariate Gaussian distribution~\cite{lee2018}.

\begin{figure}[t]
    \centering
    \includegraphics[width=\columnwidth]{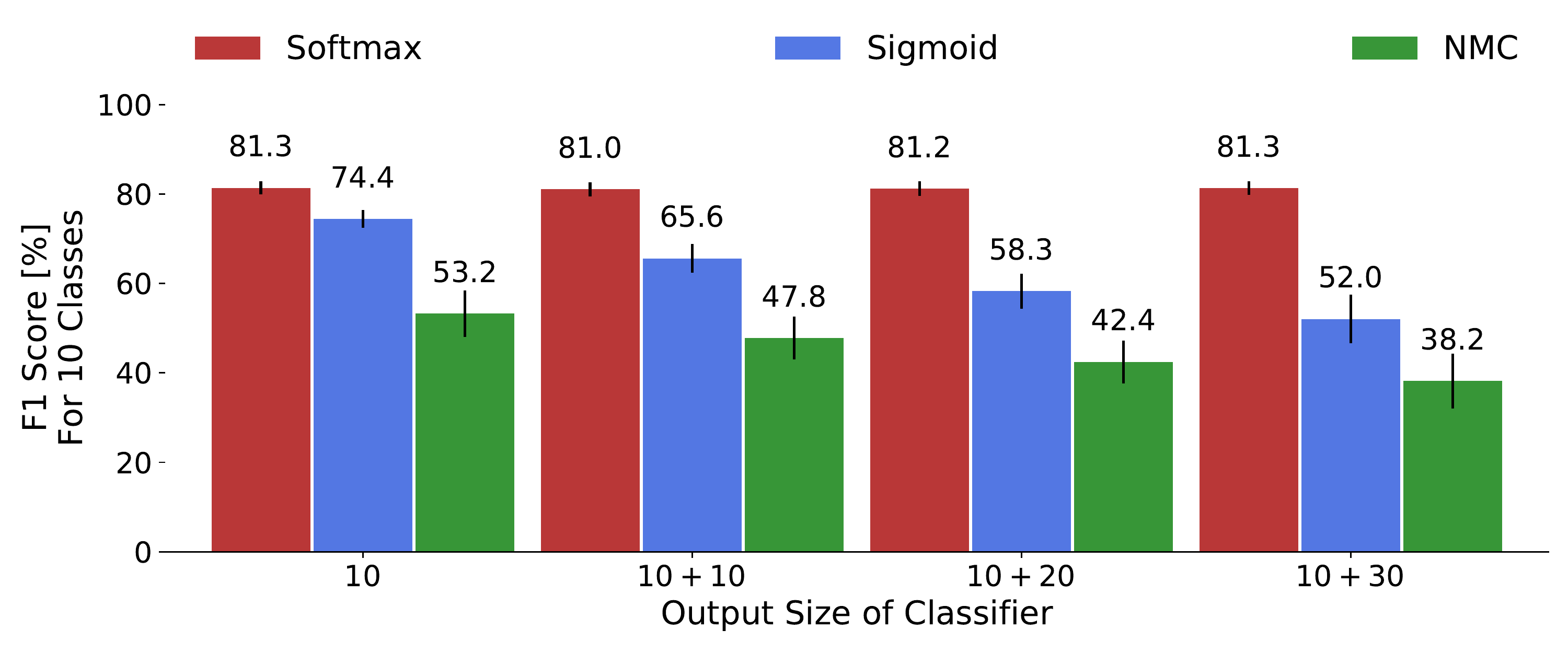}
    \caption{Impact of output layer expansion in upperbound models (base models with 10 classes).}
    \label{fig:output-size}
\end{figure}

Considering model's head expansion, \ICARL \emph{pre-allocates} the output layer: given an upperbound model, the output layer is fixed to match a larger number of classes with respect to the one expected to add at the next update. Then, a ``fake update'' is performed to push neurons of the not-yet-seen classes to zero by replaying all available known class samples, i.e. currently available classes' labels are mapped on an extended one-hot encoding.
In Fig.~\ref{fig:output-size} we take an upperbound model of 10 classes, and expand its output layer by adding $10$, $20$, and $30$ units. The first group of histograms on the left shows the base models performance; the remaining histograms show the performance after the expansion. While softmax is immune to the change, both sigmoid and NMC suffer from the expansion. We believe this is due to the intrinsic nature of the sigmoid activation function being more sensible to perturbations, which instead are smoothed out by adopting a softmax activation (thanks to its normalization).

\noindent \emph{Takeaway: We find that the best strategy is to classify via the output layer trained with a softmax activation. This further allows to expand the output layer with no extra penalty. In Fig.~\ref{fig:output-size} we show expansions at multiples of 10 units, but the same is true for any value. Thus, in \OURS we opt for expanding the output layer dynamically at each update to fit the precise number of classes required, and we classify with a model's head using a softmax activation.}

\subsection{Comparing incremental learning strategies\label{sec:results-il-strategies}}
\begin{figure}[t]
    \centering
    \includegraphics[width=\columnwidth]{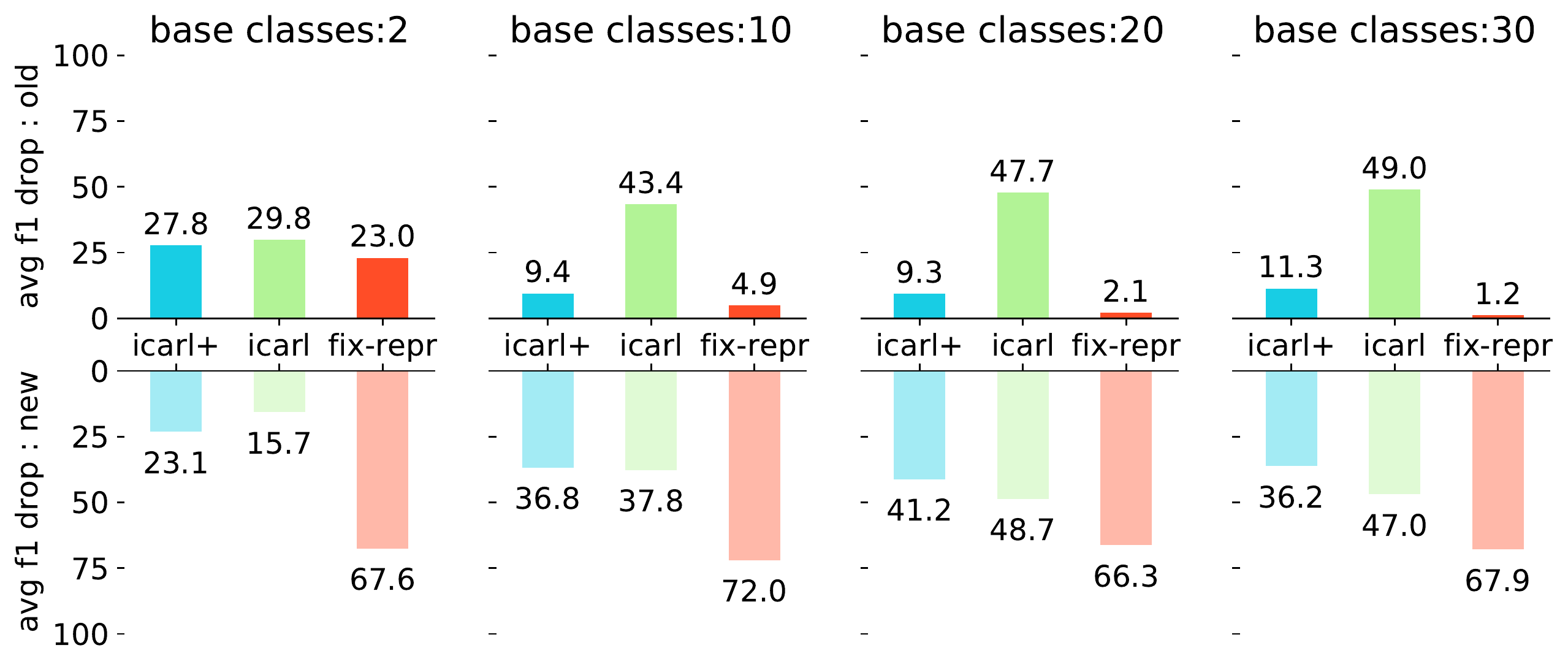}
    \caption{Comparing absolute F1 score reduction from upperbound across incremental learning methods (updates with 2 classes).
    \label{fig:increment-strategies}}
\end{figure}

Having defined the upperbound, we then compared \ICARL, \OURS, and fixed-representation performance.
We considered scenarios with base models with $2$, $10$, $20$, and $30$ classes, and performed a single update adding $2$ classes. 
Fig.~\ref{fig:increment-strategies} shows the average F1 score deviation from the upperbound, separating the evaluation for known classes (top histograms) from new classes (bottom histograms). Results corresponds to the difference between the average upperbound performance and the average performance of the incremental learning strategy across $10$ runs, i.e., the lower the value, the better the performance.

\begin{figure}[t]
    \newcommand{\w}[0]{0.24\columnwidth}
    \centering
    \subfloat[][Upperbound.]{\includegraphics[width=\w, trim=145 85 80 25, clip]{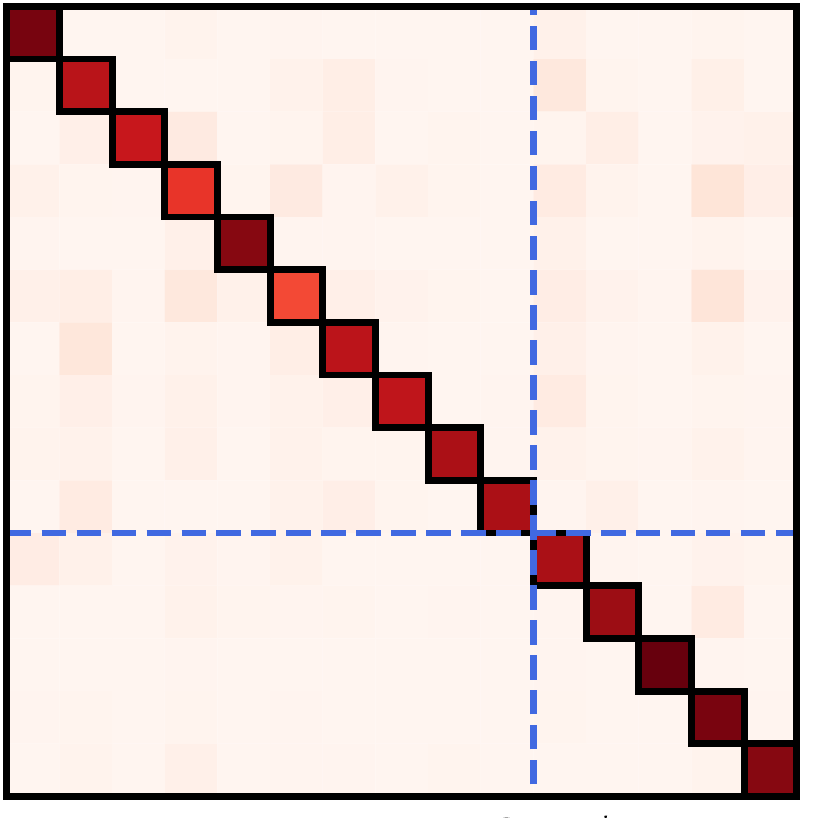}}
    \subfloat[][\OURS.]{\includegraphics[width=\w, trim=145 85 80 25, clip]{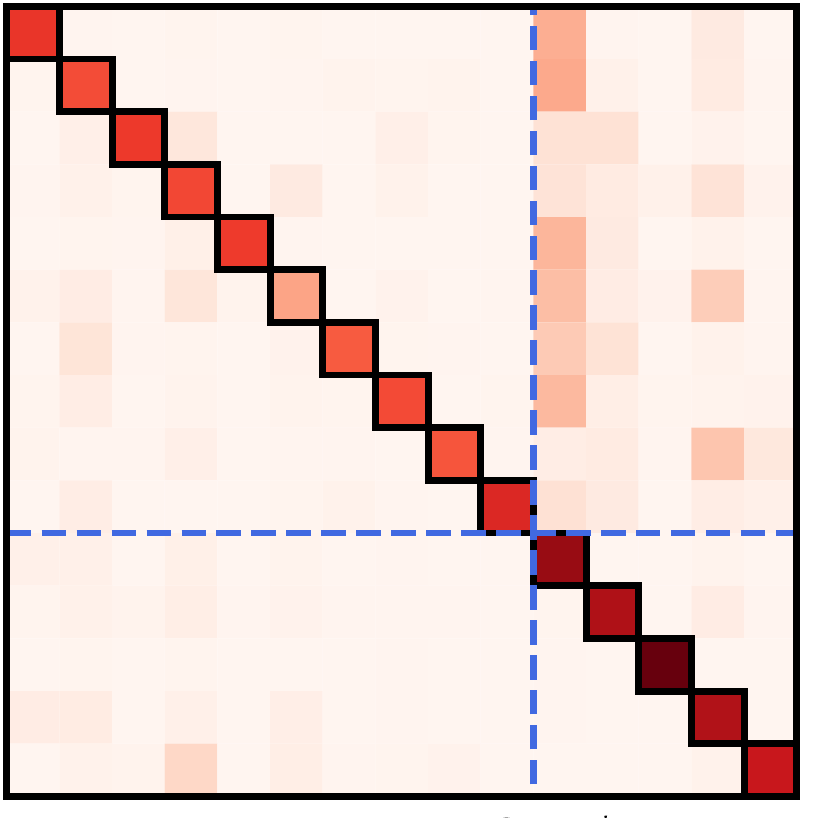}}
    \subfloat[][\ICARL.]{\includegraphics[width=\w, trim=145 85 80 25, clip]{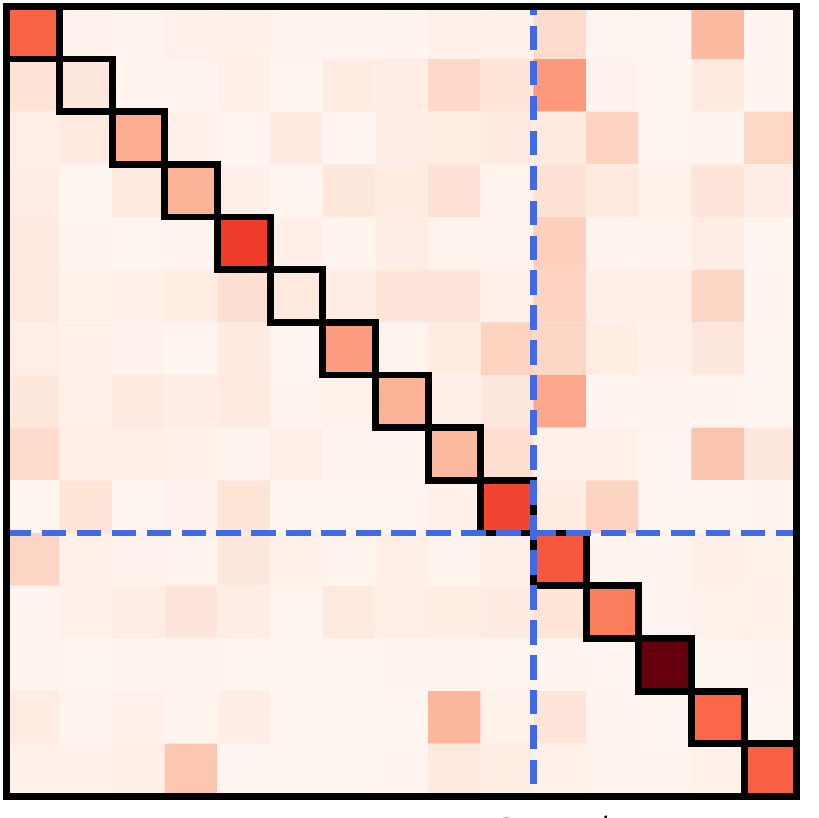}}
    \subfloat[][Fixed-repr.]{\includegraphics[width=\w, trim=145 85 80 25, clip]{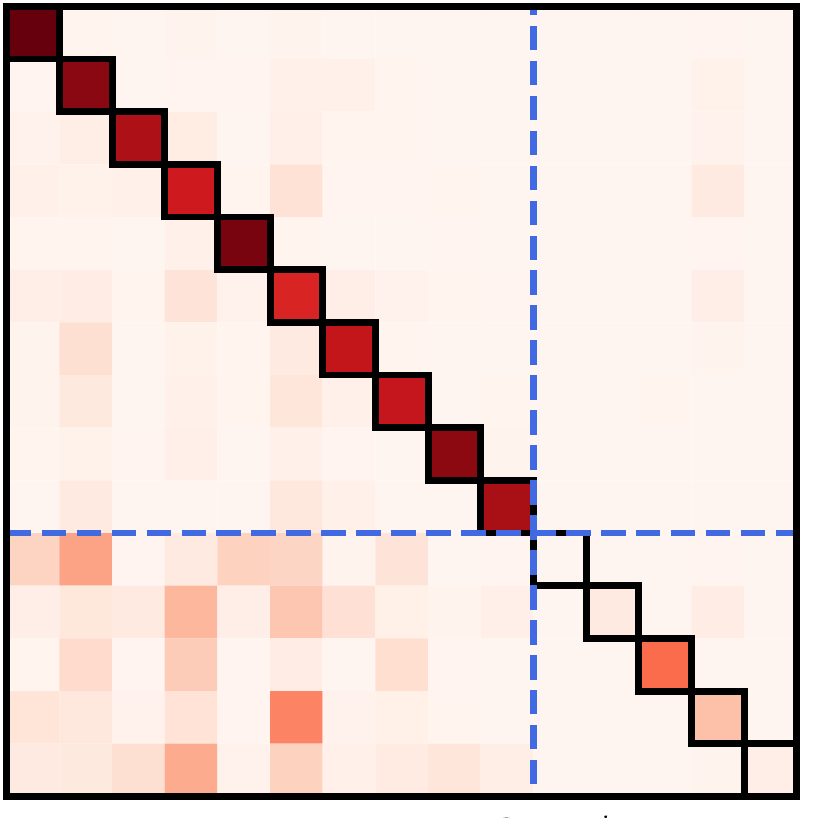}}
    \caption{Confusion matrices for one run with a base model with 10 classes, updated with $5$ new classes.
    }
    \label{fig:confusion_matrices}
\end{figure}

\ICARL suffers from more catastrophic forgetting than \OURS, especially for base models with more than $10$ classes, i.e., the practical real life scenarios. Fixed-representation instead is significantly better at preserving previous knowledge (as the model backbone is frozen), but just modifying the head does not suffice to introduce the new classes. This hints to lack of generalization of the model backbone. Overall, \OURS offers the best tradeoff, but notice how a single update with 2 classes leads to $\sim$10\% F1 accuracy drop on average for the known classes; the drop is about 3$\times$ larger for new classes. This signals an imbalance between distillation and classification loss. Fig.~\ref{fig:confusion_matrices} further renders the catastrophic forgetting across the strategies showing the confusion matrix for a single run of a base model with $10$ classes updated with $5$ classes (we use $5$ instead of $2$ just for visualization purposes). Reference lines (dashed blue) separate new and old classes. Notice that in this specific instance \OURS performance drop on new classes seems mostly due to one of the 5 new classes.

\noindent \emph{Takeaway: Performance are below expectation with respect to what reported in computer vision (e.g., average test accuracy $94.57$\%~\cite{Van2019}). We believe that integrating other mechanisms (e.g., alternative formulation of loss~\cite{lwf-TPAMI18}, re-weighting the classifier~\cite{belouadah2020-scail}, sampling strategy~\cite{borsos2020}) we could further improve \OURS, but we leave this as future work.}

\begin{figure}[t]
    \centering
    \includegraphics[width=\columnwidth]{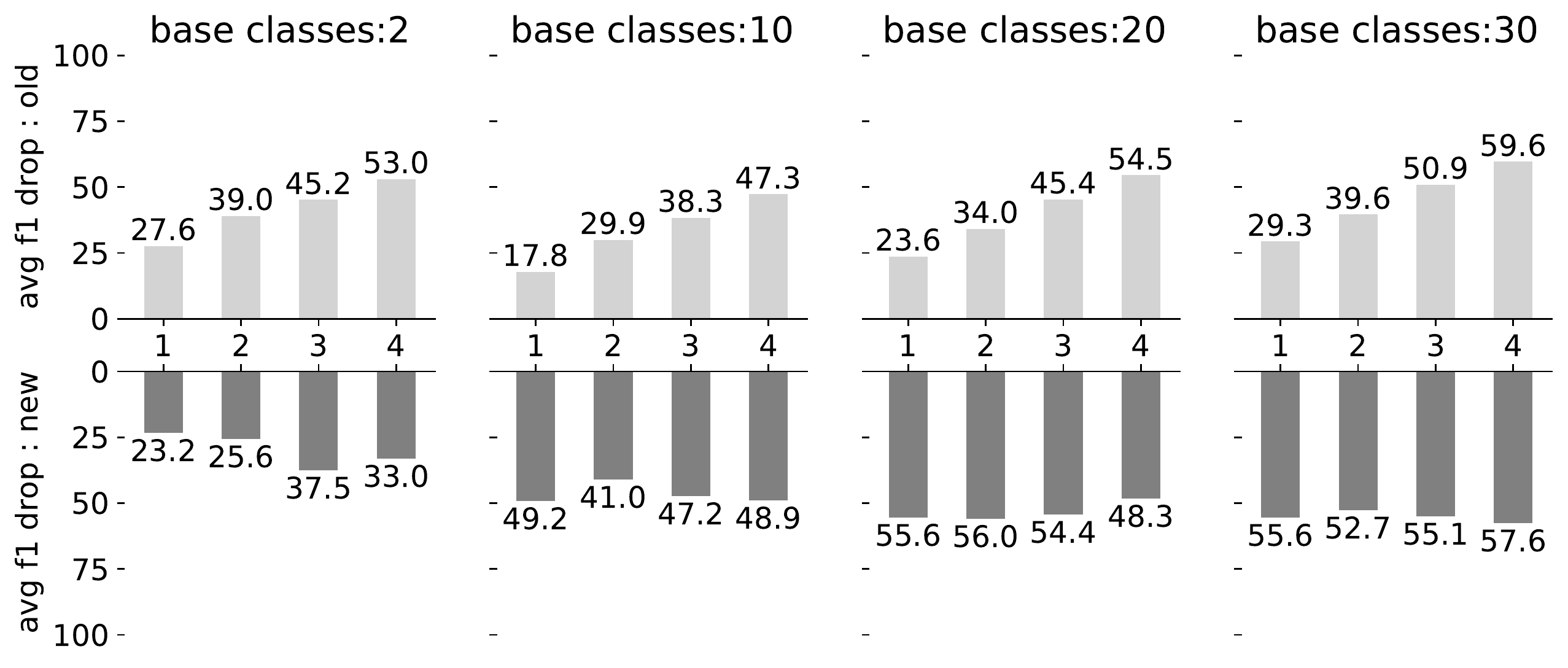}
    \caption{Absolute F1 score reduction from upperbound when performing multiple increments in sequence of 2 classes using \OURS.}
    \label{fig:multi-episodes-size-2}
\end{figure}

\begin{figure}[t]
    \centering
    \includegraphics[width=\columnwidth]{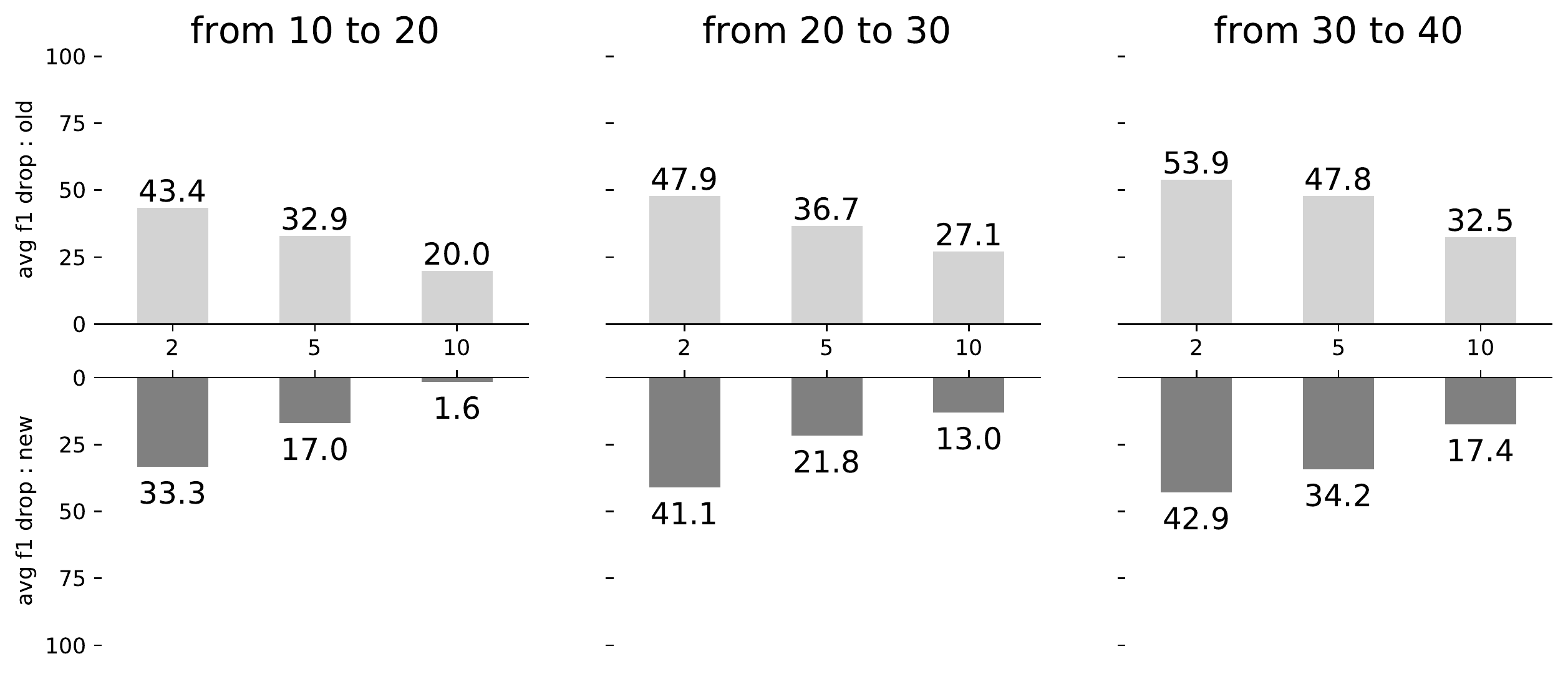}
    \caption{Absolute F1 score reduction from upperbound when performing increments of different sizes with \OURS. 
    \label{fig:multi-episodes-different-sizes}
    }    
\end{figure}

\subsection{Performing multiple episodes}

Focusing on \OURS, we now discuss how performance are affected when operating multiple updates. In Fig.~\ref{fig:multi-episodes-size-2} we show the performance drop when executing up to $4$ episodes in sequence, each adding 2 new classes each, using base models with $2$, $10$, $20$, and $30$ classes. As before, results report the average drop from the upperbound across $10$ runs. While for new classes the F1 accuracy drop is confined in the same range, catastrophic forgetting accumulates proportionally with the number of updates.

To complement the analysis, in Fig.~\ref{fig:multi-episodes-different-sizes} we compare the performance across multiple updates of different sizes. Specifically, given a model with $10$, $20$, and $30$ classes, we performed multiple updates to introduce $10$ new classes using either $5$ updates of $2$ classes each, $2$ updates of $5$ classes each, or $1$ update of $10$ classes. Two trends overlap: the larger the number of base classes, the more severe the F1 accuracy drop; the larger the number of classes in the update, the better the performance. 

\noindent \emph{Takeaway: From a practical standpoint, we expect to operate small updates, which in our analysis do not offer the best performance.
This suggests that IL should be paired with model ``checkpointing'', i.e., after a few updates a model is retrained from scratch to level its accuracy.}

\begin{figure}[t]
    \centering
    \includegraphics[width=\columnwidth]{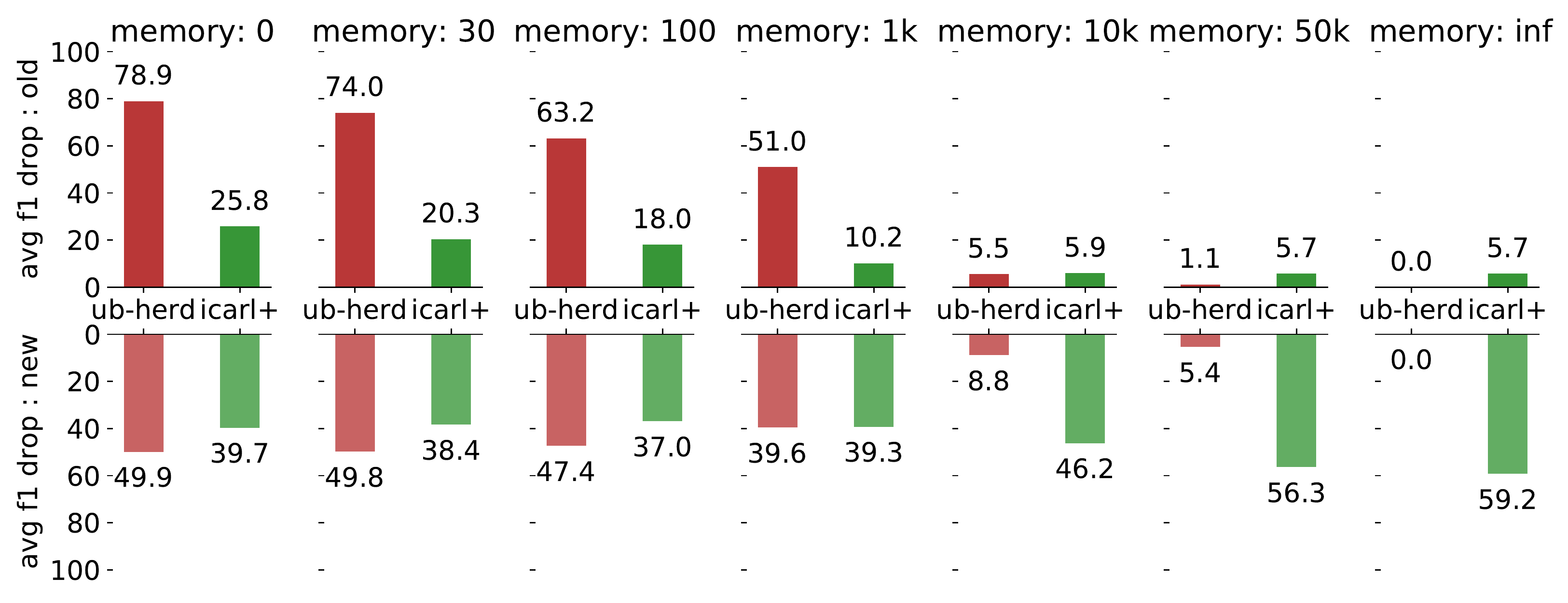}
    \caption{
    Absolute F1 score  reduction from upperbound at different memory size (base model with 10 classes, single update of 2 classes).
    }
    \label{fig:memory}
\end{figure}

\subsection{Herding selection and memory size}

We found both the herding selection and the memory size to have a key role for \OURS performance. 
To investigate this aspect, we considered a scenario with a base model of $10$ classes, and performed a single update with $2$ classes using a memory storing $100$, $1$k, $10$k, $50$k, or all exemplars. We considered also and extra reference that we named ``upperbound-herded'', i.e., given an upperbound model, we applied herding selection to identify exemplars, which we used to retrain a new upperbound from scratch---the defined upperbound-herded is a balanced downsampling strategy that we use just as an extra reference baseline.
As before, Fig.~\ref{fig:memory} shows the results of F1 accuracy drop with respect to the true upperbound, averaging results across 10 runs.

Two trends emerge. First, as expected, downsampling affects the upperbound but the herding selection is effective in selecting meaningful exemplars. Recall that the memory holds a uniform number of exemplars for each known class. In \MIRAGE a class has an average training set size of 1,800 samples, but a memory of 10k (i.e., an average 44\% reduction of training set) leads the upperbound-herded performance very close to true upperbound; it takes 10$\times$ the number of samples to match the true upperbound performance. Second, for \OURS varying the memory size trades off the distillation and classification losses, as increasing the memory size significantly reduces catastrophic forgetting (top histograms) at the cost of F1 accuracy drop for new classes (bottom histograms).

\noindent\emph{Takeaway: The memory is a key component to constrain catastrophic forgetting, but over-sizing it can be detrimental.}

\subsection{Update execution time\label{sec:results-training-time}}

\begin{figure}[t]
    \centering
    \includegraphics[width=\columnwidth]{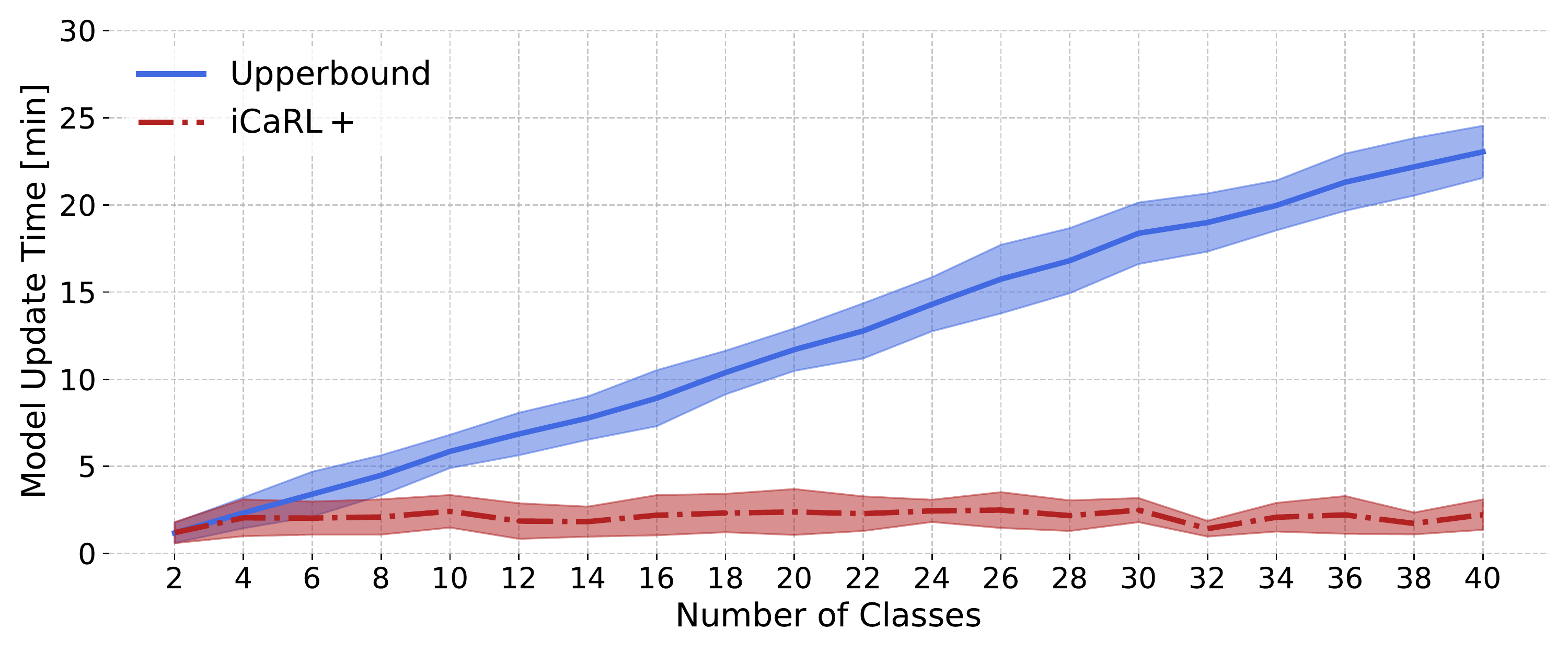}
    \caption{Model update completion time.}
    \label{fig:training-time}
\end{figure}

The killer advantage of IL is its major reduction in training time with respect to training and upperbound model. We exemplify this in Fig.~\ref{fig:training-time}: training a model with $40$ classes corresponds to $23.04$ min, against $2.23$ min to add $2$ classes to a model with $38$.

\noindent\emph{Takeaway: From the performance analysis, IL techniques seem still in their infancy. Yet, their light computational cost enables fast prototyping which is the key to unlock the model development infinite loop (Fig.~\ref{fig:dataflywheel}).}

\section{Conclusions and Future Directions}\label{sec:conclusions}

In this work we reviewed class incremental learning in DL-based traffic classification.
We dissected \ICARL's design, a state-of-the-art CIL method,
pinpointed its components, discussed alternatives design choices, and thoroughly evaluated it with the \MIRAGE dataset. We demonstrated some limitations of \ICARL's design, and we proposed a revised version \OURS based on the use of ($i$) a softmax activation function, rather than an NMC classifier, and ($ii$) a dynamic output layer expansion fitting the number of new classes, rather than a pre-sized output layer with more units that what strictly required. Although we cannot generalize our observations, no previous literature provides a similar analysis. Our analysis highlights the infancy of \OURS and similar methods given their 
lower performance on traffic classification tasks if compared to what observed for computer vision. However, it is not trivial to understand the reason behind such differences. Considering their relatively small architecture training set size, we conjecture that our models' backbone do not generalize enough, so integrating new classes still require large changes which disrupt the previously acquired knowledge. To better dissect those mechanisms, we need to contrast our analysis to more datasets, and other methods CIL methods such as~\cite{castro2018end, wu2019}.
Likewise, more robust incremental learning methodologies includes research into alternative formulation of loss functions~\cite{lwf-TPAMI18}, re-weighting the classifier and distillation loss~\cite{belouadah2020-scail}, sampling strategy~\cite{borsos2020}), and other design choices.

At a deeper level, incorporating computer vision datasets in the analysis might shed light onto \emph{domain adaptation} issues, i.e., why mechanisms proven valuable for a domain might not work as well when applied to another domain. We believe such analysis would be useful beyond the specificity of (class) incremental learning.

Overall, despite their infancy,
thanks to their massive reduction of training computational cost, we foresee a more mature set of CIL techniques as the key to materialize the vision of automation of DL-based traffic analysis systems, and closed-loop networks operation in general. However, the roadmap to such vision includes methodologies beyond CIL, such as few-shot learning and coreset identification (model classes with a very small number of samples), open-set recognition (identify new classes), etc., which so far have received limited attention by the network traffic analysis research community.

\balance
\footnotesize
\bibliographystyle{IEEEtran}
\bibliography{bibliography}

\begin{thebibliography}{10}
\providecommand{\url}[1]{#1}
\csname url@samestyle\endcsname
\providecommand{\newblock}{\relax}
\providecommand{\bibinfo}[2]{#2}
\providecommand{\BIBentrySTDinterwordspacing}{\spaceskip=0pt\relax}
\providecommand{\BIBentryALTinterwordstretchfactor}{4}
\providecommand{\BIBentryALTinterwordspacing}{\spaceskip=\fontdimen2\font plus
\BIBentryALTinterwordstretchfactor\fontdimen3\font minus
  \fontdimen4\font\relax}
\providecommand{\BIBforeignlanguage}[2]{{%
\expandafter\ifx\csname l@#1\endcsname\relax
\typeout{** WARNING: IEEEtran.bst: No hyphenation pattern has been}%
\typeout{** loaded for the language `#1'. Using the pattern for}%
\typeout{** the default language instead.}%
\else
\language=\csname l@#1\endcsname
\fi
#2}}
\providecommand{\BIBdecl}{\relax}
\BIBdecl

\bibitem{tcsurvey08comst}
T.~T. Nguyen and G.~J. Armitage, ``A survey of techniques for internet traffic
  classification using machine learning.'' \emph{IEEE Communications Surveys
  and Tutorials}, vol.~10, no. 1-4, pp. 56--76, 2008.

\bibitem{tcsurvey15survey}
P.~Velan \emph{et~al.}, ``A survey of methods for encrypted traffic
  classification and analysis,'' \emph{International Journal of Network
  Management}, vol.~25, no.~5, pp. 355--374, 2015.

\bibitem{tcsurvey18comst}
F.~{Pacheco} \emph{et~al.}, ``Towards the deployment of machine learning
  solutions in network traffic classification: A systematic survey,''
  \emph{IEEE Communications Surveys and Tutorials}, pp. 1--1, 2018.

\bibitem{bernaille06ccr}
L.~Bernaille \emph{et~al.}, ``Traffic classification on the fly,'' \emph{ACM
  SIGCOMM Computer Communication Review}, vol.~36, no.~2, pp. 23--26, 2006.

\bibitem{crotti07ccr}
M.~Crotti \emph{et~al.}, ``Traffic classification through simple statistical
  fingerprinting,'' \emph{ACM SIGCOMM Computer Communication Review}, vol.~37,
  no.~1, pp. 5--16, 2007.

\bibitem{santiago12imc}
P.~S. del Rio \emph{et~al.}, ``Wire-speed statistical classification of network
  traffic on commodity hardware,'' in \emph{Proc. ACM IMC}, 2012.

\bibitem{wang2015blackhat}
Z.~Wang, ``The applications of deep learning on traffic identification,''
  \emph{BlackHat USA}, 2015.

\bibitem{wang2017isi}
W.~Wang \emph{et~al.}, ``End-to-end encrypted traffic classification with
  one-dimensional convolution neural networks,'' in \emph{Proc. IEEE ISI},
  2017, pp. 43--48.

\bibitem{aceto2018tma}
G.~Aceto \emph{et~al.}, ``Mobile encrypted traffic classification using deep
  learning,'' in \emph{Proc. IEEE TMA}, 2018.

\bibitem{lopez2017access}
M.~Lopez-Martin \emph{et~al.}, ``Network traffic classifier with convolutional
  and recurrent neural networks for internet of things,'' \emph{IEEE Access},
  vol.~5, pp. 18\,042--18\,050, 2017.

\bibitem{aceto2019mirage}
G.~Aceto \emph{et~al.}, ``{MIRAGE}: Mobile-app traffic capture and ground-truth
  creation,'' in \emph{Proc. IEEE ICCCS}, 2019, pp. 1--8.

\bibitem{lotfollahi2020deep}
M.~Lotfollahi \emph{et~al.}, ``Deep packet: A novel approach for encrypted
  traffic classification using deep learning,'' \emph{Soft Computing}, vol.~24,
  no.~3, 2020.

\bibitem{Van2019}
G.~M. Van~de Ven and A.~S. Tolias, ``Three scenarios for continual learning,''
  in \emph{NeurIPS Continual Learning Workshop}, 2018.

\bibitem{hat-TELECOMSYST16}
V.~{Carela-Espa\~nol} \emph{et~al.}, ``A streaming flow-based technique for
  traffic classification applied to 12 + 1 years of internet traffic,''
  \emph{Telecommunication Systems}, vol.~63, p. 191–204, 2016.

\bibitem{isvm-MNA18}
G.~{Sun} \emph{et~al.}, ``Internet traffic classification based on incremental
  support vector machines,'' \emph{Mobile Networks and Applications}, vol.~23,
  p. 789–796, 2018.

\bibitem{icarl-CVPR17}
S.~{Rebuffi} \emph{et~al.}, ``{iCaRL}: Incremental classifier and
  representation learning,'' in \emph{Proc. IEEE CVPR}, 2017, pp. 5533--5542.

\bibitem{catastrophic-PSM89}
M.~{McCloskey} and N.~J. {Cohen}, ``Catastrophic interference in connectionist
  networks: The sequential learning problem,'' ser. Psychology of Learning and
  Motivation, 1989, vol.~24, pp. 109--165.

\bibitem{catastrophic-ICLR14}
Y.~{Bengio} \emph{et~al.}, ``An empirical investigation of catastrophic
  forgetting in gradient-based neural networks,'' in \emph{Proc. ICLR}, 2014.

\bibitem{batchvsincremental-IDA12}
J.~{Read} \emph{et~al.}, ``Batch-incremental versus instance-incremental
  learning in dynamic and evolving data,'' in \emph{Proc. IDA}, 2012.

\bibitem{hat-IDA09}
A.~{Bifet} and R.~{Gavald\`{a}}, ``Adaptive learning from evolving data
  streams,'' in \emph{Proc. IDA}, 2009.

\bibitem{adwin-SIAM07}
A.~{Bifet} and R.~{Gavald{\`{a}}}, ``Learning from time-changing data with
  adaptive windowing,'' in \emph{Proc. SIAM ICDM}, 2007.

\bibitem{Parker2015}
B.~Parker \emph{et~al.}, ``Incremental ensemble classifier addressing
  non-stationary fast data streams,'' in \emph{Proc. IEEE ICDMW}, vol. 2015, 01
  2015, pp. 716--723.

\bibitem{awe-KDD03}
H.~{Wang} \emph{et~al.}, ``Mining concept-drifting data streams using ensemble
  classifiers,'' in \emph{Proc. ACM KDD}, 2003.

\bibitem{inckmeans-ACISC16}
H.~R. {Loo} \emph{et~al.}, ``Online incremental learning for high bandwidth
  network traffic classification,'' \emph{Applied Computational Intelligence
  and Soft Computing}, 2016.

\bibitem{mallya2018}
A.~Mallya and S.~Lazebnik, ``{PackNet}: Adding multiple tasks to a single
  network by iterative pruning,'' in \emph{Proc. IEEE CVPR}, 2018.

\bibitem{aljundi2016}
R.~Aljundi \emph{et~al.}, ``Expert gate: Lifelong learning with a network of
  experts,'' in \emph{Proc. IEEE CVPR}, 2017, pp. 3366--3375.

\bibitem{kemker2018}
R.~Kemker and C.~Kanan, ``{FearNet}: Brain-inspired model for incremental
  learning,'' in \emph{Proc. ICLR}, 2018.

\bibitem{hayes2020}
T.~L. Hayes \emph{et~al.}, ``Remind your neural network to prevent catastrophic
  forgetting,'' in \emph{Proc. ECCV}, 2020.

\bibitem{lwf-TPAMI18}
Z.~{Li} and D.~{Hoiem}, ``Learning without forgetting,'' \emph{IEEE
  Transactions on Pattern Analysis and Machine Intelligence}, vol.~40, no.~12,
  pp. 2935--2947, 2018.

\bibitem{distillation-NIPS15}
G.~{Hinton} \emph{et~al.}, ``Distilling the knowledge in a neural network,'' in
  \emph{NIPS Deep Learning and Representation Learning Workshop}, 2015.

\bibitem{belouadah2020}
E.~Belouadah \emph{et~al.}, ``A comprehensive study of class incremental
  learning algorithms for visual tasks,'' \emph{Neural Networks}, 2020.

\bibitem{Shin2017}
\BIBentryALTinterwordspacing
H.~Shin \emph{et~al.}, ``Continual learning with deep generative replay,''
  \emph{CoRR}, vol. abs/1705.08690, 2017. [Online]. Available:
  \url{http://arxiv.org/abs/1705.08690}
\BIBentrySTDinterwordspacing

\bibitem{multitasklearn-ICCCN20}
S.~{Rezaei} and X.~{Liu}, ``Multitask learning for network traffic
  classification,'' in \emph{Proc. ICCCN}, 2020.

\bibitem{neverending-COMM18}
T.~{Mitchell} \emph{et~al.}, ``Never-ending learning,'' \emph{Commun. ACM},
  vol.~61, no.~5, p. 103–115, 2018.

\bibitem{fewshot-IEEEAccess19}
H.~Sun \emph{et~al.}, ``Common knowledge based and one-shot learning enabled
  multi-task traffic classification,'' \emph{IEEE Access}, vol.~7, pp.
  39\,485--39\,495, 2019.

\bibitem{herding-ICML09}
M.~{Welling}, ``Herding dynamical weights to learn,'' in \emph{Proc. ICML},
  2009.

\bibitem{nmc-TPAM13}
T.~{Mensink} \emph{et~al.}, ``Distance-based image classification: Generalizing
  to new classes at near-zero cost,'' \emph{IEEE Transactions on Pattern
  Analysis and Machine Intelligence}, vol.~35, no.~11, pp. 2624--2637, 2013.

\bibitem{nmc-CVPR14}
M.~{Ristin} \emph{et~al.}, ``Incremental learning of {NCM} forests for
  large-scale image classification,'' in \emph{Proc. IEEE CVPR}, 2014.

\bibitem{transflearn-INFOCOM20}
J.~{Zhang} \emph{et~al.}, ``Autonomous unknown-application filtering and
  labeling for {DL-based} traffic classifier update,'' in \emph{Proc. IEEE
  INFOCOM}, 2020.

\bibitem{castro2018end}
F.~M. Castro \emph{et~al.}, ``End-to-end incremental learning,'' in \emph{Proc.
  ECCV}, 2018, pp. 233--248.

\bibitem{wu2019}
Y.~Wu \emph{et~al.}, ``Large scale incremental learning,'' in \emph{Proc. IEEE
  CVPR}, 2019, pp. 374--382.

\bibitem{mcpa-TMA19}
G.~{Tangari} \emph{et~al.}, ``Tackling mobile traffic critical path analysis
  with passive and active measurements,'' in \emph{Proc. IEEE TMA}, 2019.

\bibitem{lee2018}
K.~Lee \emph{et~al.}, ``A simple unified framework for detecting
  out-of-distribution samples and adversarial attacks,'' in \emph{Proc.
  NeurIPS}, 2018.

\bibitem{belouadah2020-scail}
E.~Belouadah and A.~Popescu, ``{ScaIL}: Classifier weights scaling for class
  incremental learning,'' in \emph{Proc. IEEE/CVF WACV}, 2020.

\bibitem{borsos2020}
Z.~Borsos \emph{et~al.}, ``Coresets via bilevel optimization for continual
  learning and streaming,'' in \emph{Proc. NeurIPS}, 2020.

\end{thebibliography}

\end{document}